# Collisions and compositional variability in chondrule-forming events

Emmanuel Jacquet[1]

[1]Institut de Minéralogie, de Physique des Matériaux et de Cosmochimie (IMPMC), Muséum national d'Histoire naturelle, Sorbonne Université, CNRS ; CP52, 57 rue Cuvier, 75005 Paris, France.

E-mail: emmanuel.jacquet@mnhn.fr

*Abstract*

Compound chondrules, i.e. chondrules fused together, make a powerful probe of the density and compositional diversity in chondrule-forming environments, but their abundance among the dominating porphyritic textures may have been drastically underestimated. I report herein microscopic observations and LA-ICP-MS analyses of lobate chondrules in the CO3 chondrites Miller Range 07193 and 07342. Lobes in a given chondrule show correlated volatile and moderately volatile element abundances but refractory element concentrations are essentially independent. This indicates that they formed by the collision of preexisting droplets whose refractory elements behaved in closed system, while their more volatile elements were buffered by the same gaseous medium. The presence of lobes would otherwise be difficult to explain, as surface tension should have rapidly imposed a spherical shape at the temperature peak. In fact, since most chondrules across chondrite groups are nonspherical, a majority are probably compounds variously relaxed toward sphericity. The lack of correlation of refractory elements between conjoined compound chondrule components is inconsistent with derivation of chondrules from the disruption of homogenized melt bodies as in impact scenarios and evokes rather the melting of independent mm-size nebular aggregates. Yet a "nebular" setting for chondrule formation would need to involve not only increased solid concentration, e.g. by settling to the midplane, but also a boost in relative velocities between droplets during chondrule-forming events to account for observed compound chondrule frequencies .

## 1. Introduction

Chondrules, the millimeter-sized igneous spheroids that make up the bulk of primitive meteorites, or chondrites, have not surrendered the secret of their origin yet (e.g. Russell et al. (2018) ). Four decades ago, building on the experience of lunar rocks and terrestrial impact craters, Taylor et al. (1983) had argued against "planetary scenarios" such as impacts or volcanism based on the lack of correlated shock effects or the primitive and the variable bulk composition of chondrules. With the discovery of relict grains (e.g. Jones (1996), Marrocchi et al. (2019), Nagahara (1981), Rambaldi et al. (1983), Schrader et al. (2018) ), the focus of most chondrule students turned to "nebular scenarios" envisioning the "flash" melting of primordial dust aggregates freely floating in the protoplanetary disk, e.g. during passage of shock waves (e.g. Desch et al. (2005) ) or lightning discharges (e.g. Desch and Cuzzi (2000), Johansen and Okuzumi (2018) ). However, in the 2000s, evidence resurfaced that chondrule-forming environments were much richer in solids than expected in "average" protoplanetary disks. The FeO contents of chondrule ferromagnesian silicates, at variance with the reducing conditions of a solar gas ( Grossman et al. (2012) ), suggest solid/gas ratios one to three orders of magnitude above solar (e.g. Schrader et al. (2011), Tenner et al. (2015) ). The retention of moderately volatile sodium may call for even higher densities (e.g. Alexander et al. 2008), unless cooling was very rapid (e.g. Wasson and Rubin 2003). Thus, planetary scenarios such as impact jetting ( Johnson et al. (2015) ) or "splashing" of preexisting magma oceans (Asphaug et al. 2011; Sanders and Scott (2012) ) have regained the favor of many researchers, even though difficulties remain as to the apparently primitive composition of chondrule precursors ( Jacquet and Marrocchi (2017), Jones and Schilk (2009), Misawa and Nakamura (1988) ) or their narrow size distribution ( Jacquet (2014) ).

Crucial to the "planetary" vs. "nebular" debate would be to know how chondrule compositions varied in *single* chondrule-forming events. Schematically, a "planetary" (impact) scenario would locally make chondrules out of a single melt body and thus would predict uniform compositions in a given plume ( Jacquet (2014), Taylor et al. (1983) ). In contrast a "nebular" scenario, where chondrule precursors are stochastic aggregates of nebular solids, would predict no correlation between them. Indeed, the proportions of different varieties of such nebular solids would vary from neighbor to neighbor; some may e.g. have incorporated less and other more refractory inclusions. Examples of the latter may include chondrules with volatility-fractionated Rare Earth Element (REE) patterns ( Jacquet and Marrocchi (2017) ; Misawa and Nakamura (1988) ; Jones and Schilk (2009); Zhang et al. (2020)).

Unfortunately, the mere coexistence of two chondrules in some meteorite does not prove their co-formation, as any given chondrite probably accreted chondrules from different sources, if only in terms of redox conditions, viz. type I (reduced) vs. type II (oxidized) chondrules. However, *compound chondrules*, that is pairs or multiplets of chondrules fused together ( Gooding and Keil (1981), Lux et al. (1981), Wasson et al. (1995) ; Akaki and Nakamura (2005) ) may offer a glimpse into individual chondrule-forming environment diversity. Indeed, although Hubbard (2015) advocated cold formation, the consensus is that they fused together while being partially molten (e.g. Arakawa and Nakamoto (2016), Gooding and Keil (1981) ; Wasson et al. (1995)). Thus, the component chondrules must have collided during the high-temperature event that melted them. In the literature, bulk analyses of compound chondrule components by electron microprobe (defocused beam or averaged spots) have been reported by Lux et al. (1981) and Akaki and Nakamura (2005) but were not specifically compared, even though Wasson et al. (1995) did correlate the FeO contents of their ferromagnesian silicates.

Literature data suggest compound chondrules are rather uncommon. Gooding and Keil (1981) found a proportion of 4 % from disaggregation experiments of ordinary chondrites, a result that seems consistent with the 2.4 % frequency found by Wasson et al. (1995) in their extensive thin section survey if 2D sectioning effects are allowed for ( Ciesla et al. (2004) ). The latter number compares reasonably with the 1.6 % found in sections of the Allende and Axtell CV3 carbonaceous chondrites by Akaki and Nakamura (2005) and the 1 % found for a set of 50 Antarctic unequilibrated carbonaceous and ordinary chondrites studied by Sekiya and Nakamura (1996).

Yet the statistics of compound chondrules may not be as well-established as this consistency in the literature suggests. It has long been known that nonporphyritic chondrules (barred, radial, cryptocrystalline) are overrepresented in compound chondrules (e.g. Gooding and Keil 1981; Wasson et al. 1995). In fact, only *one* out of 83 compound chondrules studied by Wasson et al. (1995) contained only porphyritic textures, although such textures dominate in the general chondrule population (e.g. Rubin (2010) ). This may betray an unavoidable recognition bias.

Indeed, in optical microscopy, it is much easier to recognize a chondrule as compound when at least one of the components is nonporphyritic, because of textural contrast and/or the deformation of the (undercooled) secondary chondrule against the rigid (already crystallizing) primary chondrule. Two colliding partially molten porphyritic chondrules would merely combine their crystals and melts, with little textural contrast and no definable primary/secondary boundary—indeed no definable primary or secondary chondrule at all. Akaki and Nakamura (2005) were the first to recognize and define such *blurred* compound chondrules (i.e. those lacking a circular arc at the interface) in CV chondrites. Although those only increased their compound chondrule count by a third, there may be more, less easily recognizable, as surface tension may have more or less restored sphericity.

In CO chondrites in particular, Rubin and Wasson (2005) found that more than 60 % of chondrules had lobate shapes. So one may entertain the thought that many are hitherto overlooked compound objects. I have thus set to examine microscopically such objects in CO chondrites, to assess their potential compound nature, and to measure the bulk compositions of their individual lobes using *in situ* laser ablation inductively coupled plasma mass spectrometry (LA-ICP-MS). The comparison of the lobe compositions (from this study and the literature) will inform the question of the nature of chondrule precursors and that of the chondrule-forming environments. To assess the full range of chondrule compositions, I also analyzed a few mesostasis-(and thus Al-)rich chondrules in the same chondrite, hoping to add more volatility-fractionated patterns to those of Jacquet and Marrocchi (2017) mentioned above.

## 2. Samples and methods

I studied the two thick sections MIL 07342,9 and MIL 07193,19 from CO3 chondrites Miller Range (MIL) 07342 and MIL 07193, from NASA's Antarctic Meteorite collection. While Sears (2016) considered pairing between the two likely, they are listed in separate pairing groups in the said database. Objects identified in MIL 07342 and MIL 07193 are given the prefixes "M" (as in M1, M2, etc.) and "Mi" (as in Mi1, Mi2, etc.), respectively.

The objects of interest were examined in scanning electron microscopy (SEM) with a JEOL JSM-840A with EDAX EDS detector at the Muséum national d'Histoire naturelle. X-ray maps (more precisely the false-color overlays such as those shown later in the paper) allowed apparent mineral modes of chondrules to be calculated using the JMicrovision software (N. Roduit; https//jmicrovision.github.io; last accessed in April 2020) which also allowed 2D geometrical measurements. Minor and major element concentrations of documented chondrules were obtained with a Cameca SX-100 electron microprobe (EMP) at the Centre de Microanalyse de Paris VI (CAMPARIS).

Trace element analyses of chondrules were performed by LA-ICP-MS at the University of Montpellier II. The laser ablation system was a GeoLas Q$^+$ platform with an Excimer CompEx 102 laser and was coupled to a ThermoFinnigan Element XR mass spectrometer. The ICP-MS was operated at 1350 W and tuned daily to produce maximum sensitivity for the medium and high masses, while keeping the oxide production rate low ($^{248}$ThO/$^{232}$Th ≤ 1%). Ablations were performed in pure He-atmosphere (0.65 ± 0.05 L/min) mixed before entering the torch with a flow of Ar (≈ 1.00 ± 0.05 L/min). Laser ablation conditions were: fluences ca. 12 J/cm² with pulse frequencies between 5 and 10 Hz were used and spot sizes of 26-163 μm. With such energy fluences, depth speed for silicates is about 1 μm/s. Each analysis consisted of 4 min of background analyses (laser off) and 40 s of ablation (laser on). Data reduction was carried out using the GLITTER software ( Griffin et al. (2008) ). Internal standard was Si (Ca for augite), known from EMP analyses. The NIST 612 glass ( Pearce et al. (1997) ) was used as an external standard. Most of the analyses (especially all those of compound chondrules) were bulk analyses (Si being reconstructed from EMP analyses and measured modes), but in a few chondrules, mineral (olivine, pyroxene and/or mesostasis)-scale analyses were performed. A total of 36 chondrules were analysed. Examination of the time-resolved GLITTER signal allowed exclusion of contaminating phases if those appeared during ablation; the absence of such phases from the start was checked by SEM imaging of ablation craters (compared with prior images) and comparison with EMP data. For each object, an arithmetic averaging was used to calculate mean concentrations (except for silicates, where a geometric averaging was used to further minimize potential contamination for incompatible elements). The error was identified with the instrumental error for a single analysis, and with the standard error of the mean when several were averaged, so as to also account for nonrepresentative 2D sectioning effects (neglecting textural correlations between LA-ICP-MS spots). CI chondrite concentrations used for normalization were taken from Lodders (2003).

# 3. Results

### Textures

Pictures of all chondrules studied are presented in the Electronic Annex. Most chondrules in MIL 07193 and MIL 07342 are porphyritic, ferromagnesian chondrules of type I (with Fe/(Mg+Fe) ranging from 0.5 to 5.7 mol% in olivine). Except for pyroxene-rich chondrules (or rather chondrule sections), where olivine chadacrysts are scattered fairly evenly throughout their volume, olivine tend to concentrate near the chondrule interior, and (augite-overgrown) enstatite near the margin as commonly seen in (especially carbonaceous) chondrites (e.g. Friend et al. (2016), Libourel et al. (2006) ). Chondrule mesostases are usually microcrystalline, with augite microlites recognizable in an Al-rich (~feldspathic) background. Opaque minerals comprise Fe-Ni metal (better preserved in our MIL 07193 section), sulfides, magnetite and finer-grained oxides (with molar O/Fe around 2 according to the EDS) presumably attributable to terrestrial weathering. When few in number (and coarse in size), opaque grains tend to be in the chondrule periphery, as observed e.g. for sulfides in CV chondrites ( Marrocchi and Libourel (2013) ) or metal in CR chondrites ( Connolly et al. (2001), Jacquet et al. (2013), Lee et al. (1992), Wasson and Rubin (2010) ). Petrographic data for all studied chondrules are given in the Electronic Annex. Trace element *mineral* analyses were obtained for 4 porphyritic chondrules.

Mesostasis modes average 9 vol% (including Ca-rich pyroxene). I selected 9 type I mesostasis-rich chondrules exceeding this value for analysis, some depicted in Fig. 1. Five are barred olivine (± pyroxene) chondrules and four aluminium-rich chondrules (ARC) which may contain visible spinel and/or anorthite. I also analyzed a (barely) type II mesostasis-rich PP chondrule, Mi33 (Fig. 2a) conspicuous for its presence of silica (also seen in two type I chondrules not included in this study).

Chondrules are rarely strictly spherical. They may be merely elongated in shape, but many show distinct constrictions and lobes (two or more), as depicted in Fig. 3 and 4. As observed by Rubin and Wasson (2005), there is a continuum of shapes, between well-defined necks (e.g. M6; Fig. 3a) and mere bumps on largely spheroidal objects and a continuum of size ratios, from subequal size-components (e.g. Mi11; Fig. 4a) to small adhesions (e.g. M31; Fig. 3c). Most of the time, the lobes show similar ferromagnesian porphyritic textures, but contacts between ferromagnesian and aluminium-rich lithologies occur (e.g. Fig. 5). For ferromagnesian mineralogies, if olivine is abundant, it tends to form distinct clusters in the individual lobes, with the "neck" being dominated by enstatite, and occasionally relatively coarse opaque grains; that is, the typical mineralogical zoning of type I chondrules mentioned in the first paragraph exists at the scale of individual lobes. Such necks can also be seen in Renazzo chondrules studied by Ebel et al. (2008). I selected 18 type I multilobate chondrules—plus a type II one (Mi37), excluded from the correlation study—for analysis. I named individual lobes, or "components", with letter suffixes, with "a" denoting the largest (or "main") component (as for Mi40a in Mi40), and "b", "c" etc. (if applicable) the others.

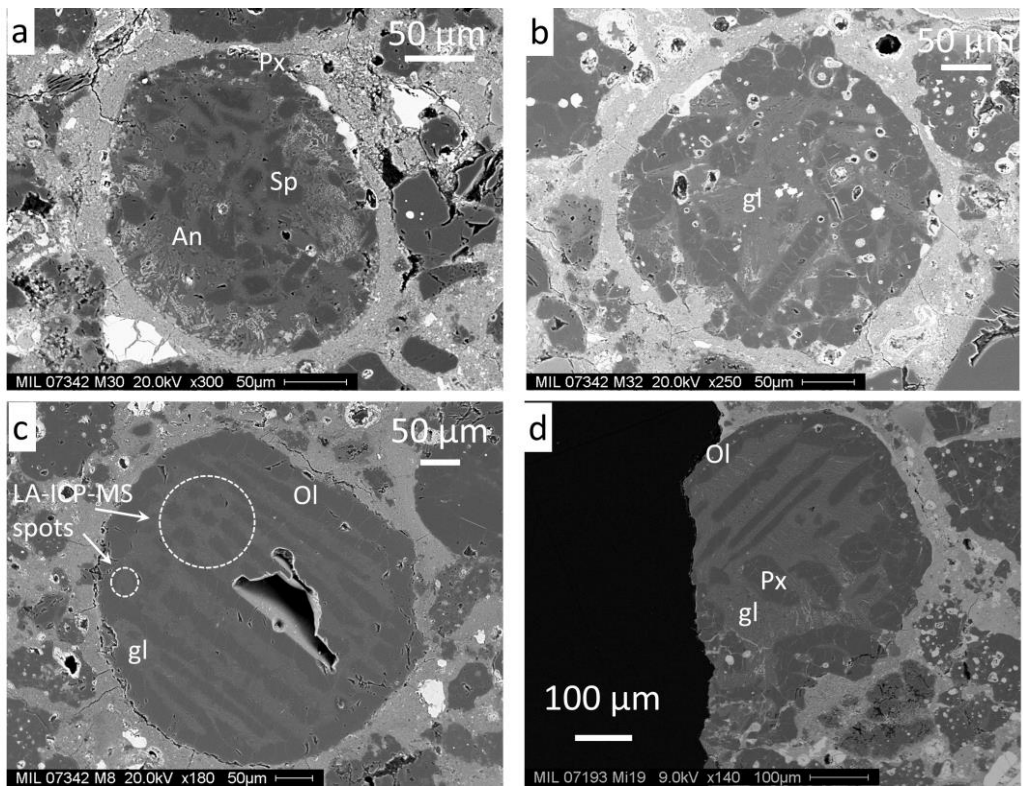

**Figure 1**: Back-scattered electron images of mesostasis-rich chondrules. a) The ovoid aluminum-rich chondrule M30 contains anhedral spinel crystals embedded in a mesostasis of anorthite laths and interstitial silica in its interior, with an outer margin dominated by enstatite and augite, along with alteration products. b) Porphyritic pyroxene chondrule M32 contains abundant microcrystalline mesostasis, enstatite laths poikilitically enclosing scarce, small olivine chadacrysts. Small (<10 μm) grains of metal, sulfides, magnetite and other oxides occur. c) Barred olivine chondrule M8. The mesostasis and bulk analyses, which gave the contrasting results mentioned in the "Result" section, are located. Olivine bars give way to equidimensional olivine grains in the vicinity of these spots (upper left). d) Chondrule Mi19 (interrupted on the left by the section's edge) has a partly digested barred olivine texture in its upper half while sometimes hopper textured enstatite crystals abound in the lower

half, with frequent augite overgrowths. The mesostasis is microcrystalline. Ol = olivine, Px = low-Ca pyroxene, Cpx = Ca-rich pyroxene. Sp = spinel, An = anorthite, gl = mesostasis.

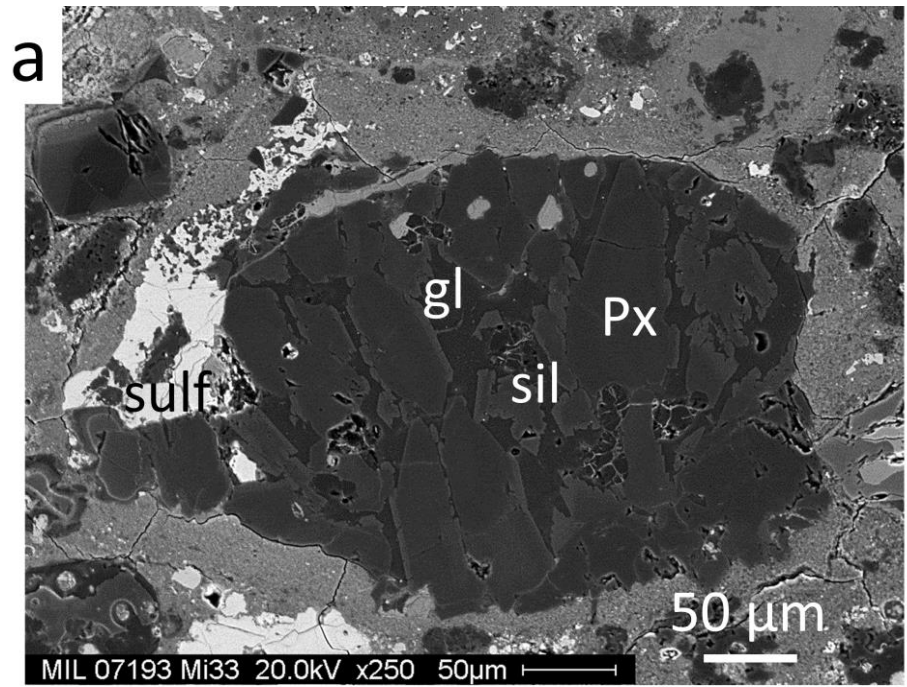

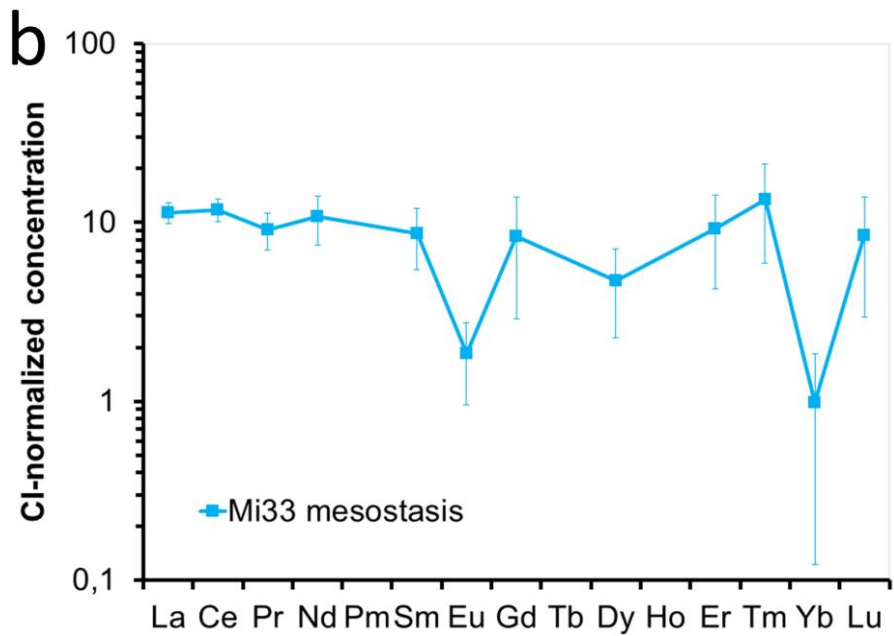

**Figure 2**: Type II porphyritic pyroxene chondrule Mi33. a) Back-scattered electron image. Low-Ca pyroxene (Px; $Fs_{11}$) overgrown by augite are embedded in largely glassy, Ca-poor (1,2 wt% Ca) mesostasis (gl). Silica (sil, black), partly altered occurs near or within it. On the left margin of this elongated chondrule, sulfide (sulf) is intergrown with silicates. b) REE pattern of mesostasis. Negative anomalies in Eu and Yb are prominent. Error bars are one standard deviation.

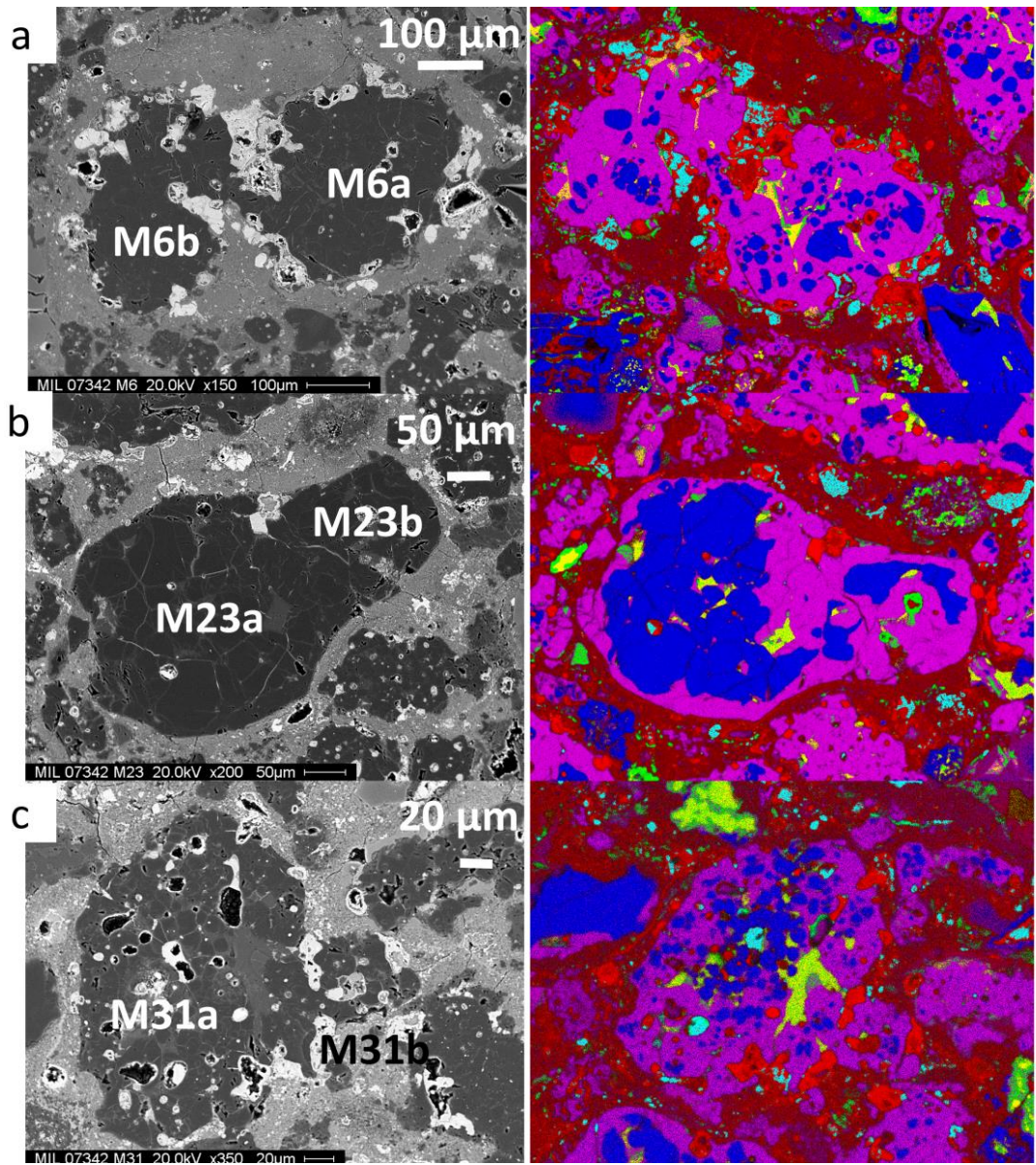

**Figure 3**: Lobate chondrules from MIL 07342. a) M6. b) M23. c) M31. Left panels are back-scattered electron images, where the names of the individual lobes are indicated. Right panels are composite X-ray maps (mind the ~30° clockwise rotation) with blue = Mg (here mostly olivine), pink = Si (mostly enstatite), green = Ca (mostly augite), yellow = Al (mostly mesostasis), cyan = S (mostly sulfides), red = Fe (mostly metal, magnetite and other oxides).

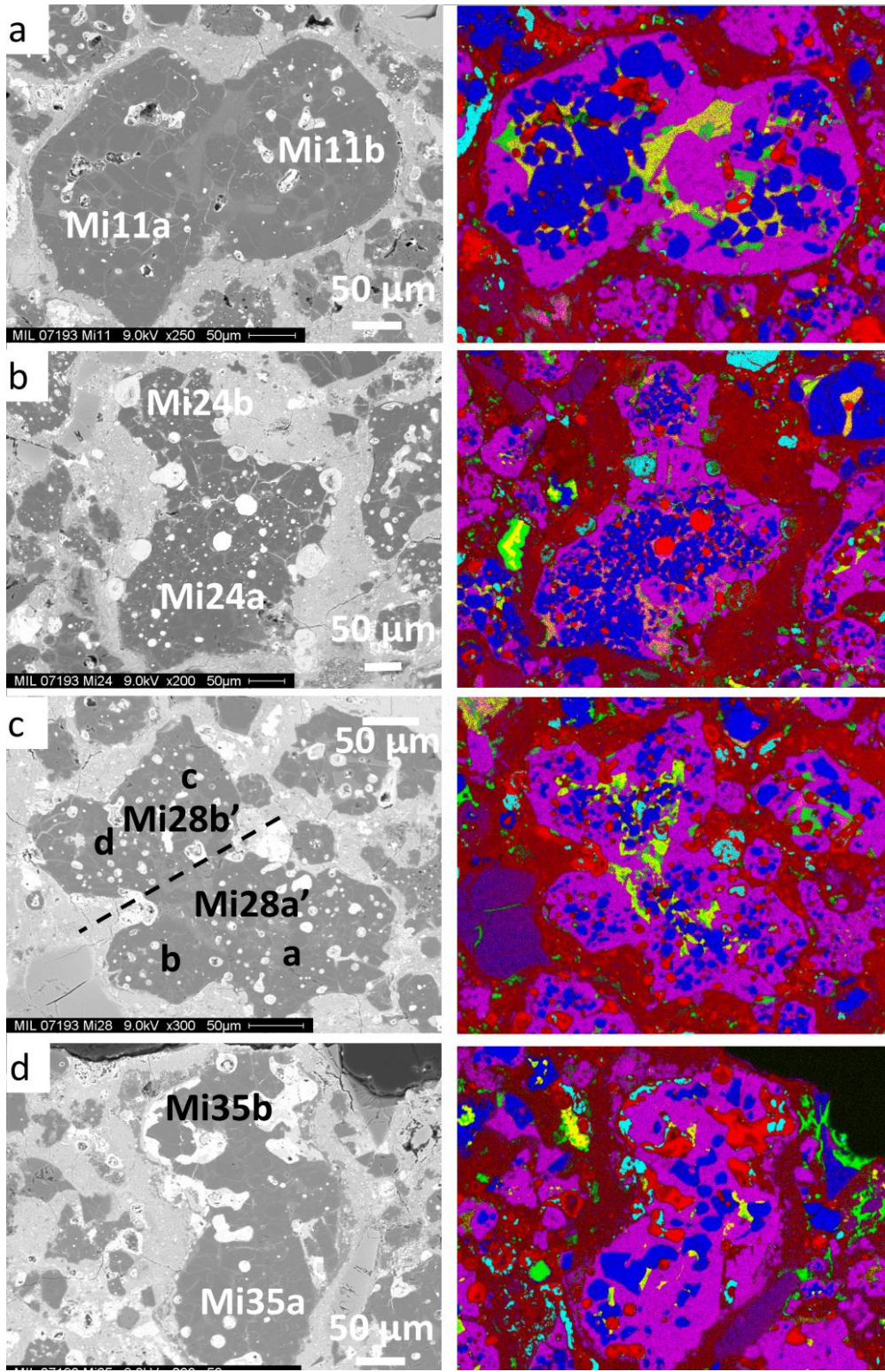

**Figure 4**: Same as Fig. 3, for MIL 07193 compound chondrules. a) Mi11. b) Mi24. c) Mi28. 4 lobes are distinguished (a,b,c,d), but I also depict a more conservative twofold partition in a' and b' (adopted for the interlobe correlation study) which may indicate the final collision (after the a-b and c-d collisions). d) Mi35.

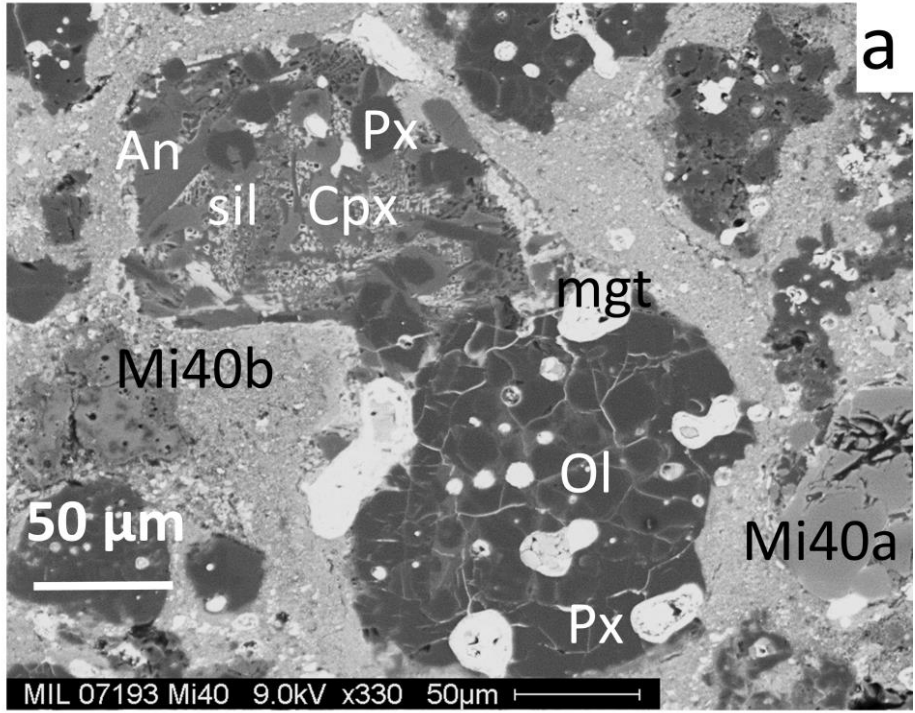

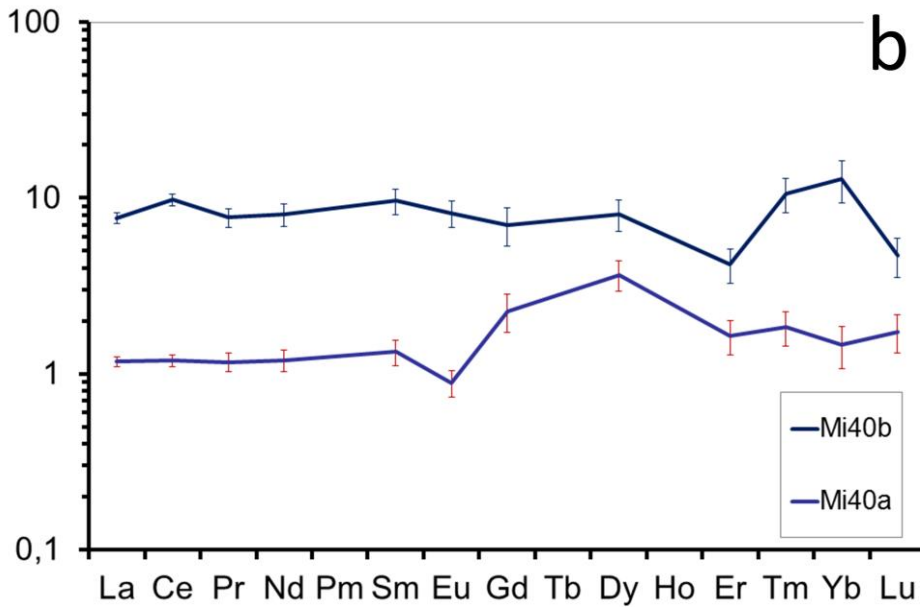

**Figure 5**: Compound chondrule Mi40. a) Back-scattered electron image. The lower right lobe (Mi40a) has a typical POP texture, with rounded olivine phenocrysts clustering near the center and enstatite dominating near the margin. The upper left lobe (Mi40b) has a pyroxene-anorthite-silica mineralogy comparable to M30 (Fig. 1a), except for the absence of spinel. b) Bulk REE patterns of the two components. Mi40a's is fairly flat, if somewhat HREE enriched, around $2 \times$ CI while Mi40b has a subdued "modified" group II pattern (in the sense of Hiyagon et al. (2011) ), around $10 \times$ CI. Error bars are one standard deviation.

Chemistry

All compositional data, individual or averaged, are available in the Electronic Annex. If we use the main lobes of the lobate chondrules as a proxy for the general chondrule population (Fig. 6), we see that chondrules in the studied CO chondrites are depleted in moderately and volatile elements as a generally smooth function of the Lodders (2003) half-condensation temperature $T_{50}$ (below 1300 K), down to one order of magnitude below refractory elements around 700 K. The siderophile elements Co, Cu, Ni are also depleted in spite of cosmochemically belonging to the "main component" ($T_{50}$ ~ 1350 K). Except for this latter depletion, the chondrules roughly follow the trend of the bulk chondrite (Braukmüller et al. 2018; Fig. 6), if with a somewhat steeper slope. This is comparable to CM chondrule analyses reported by van Kooten et al. (2019).

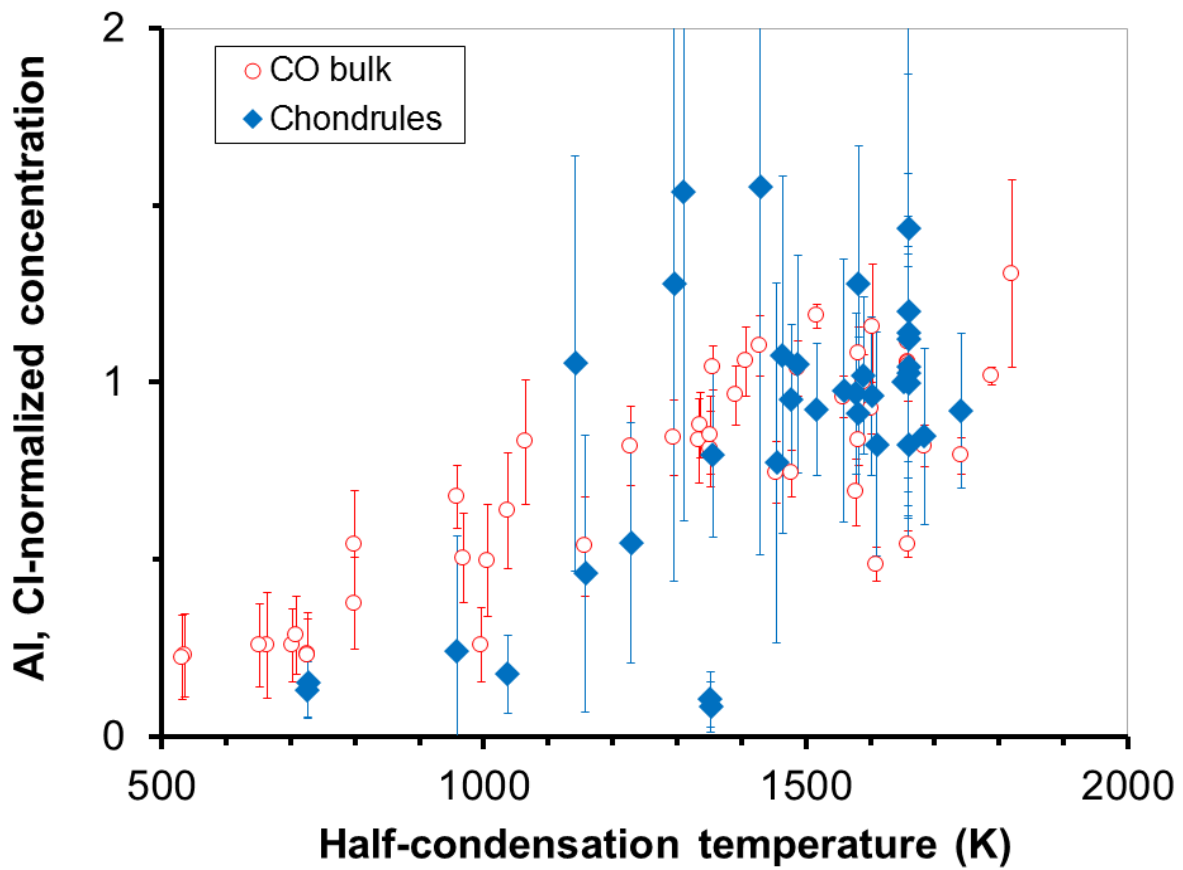

**Figure 6**: Average CI- and Al-normalized concentrations of measured elements in chondrule main lobes, compared to CO average bulk composition after Braukmüller et al. (2018), as a function of volatility under solar conditions (Lodders 2003). Error bars are one standard deviation. For the bulk CO data, the error bars do not correspond to analytical errors, but to the standard deviation of the CO, CM, CV groups (each given equal weight), as a proxy for parent body variability (see Discussion).

REE patterns of the analyzed chondrules are generally flat, but some volatility-fractionated patterns occur. Defining $Tm^*_N = (Er_N \times Yb_N)^{1/2}$ where "N" denotes CI-normalization, we have $Tm/Tm^*$ (=$Tm_N/Tm^*_N$) ranging between 0.9 and 3.9. Patterns describable as diluted group II (with heavy REE depletion and positive Tm anomalies) or modified group II (with additional Yb enhancement; Hiyagon et al. (2011) ) are seen for 5 out of the 9 mesostasis-rich type I chondrules analysed (Fig. 7, where either the bulk or the mesostases, which should be the main carriers of the incompatible REE, were analyzed). This is more than the 4 out of 26 representative (not necessarily mesostasis-rich) chondrules found by Jacquet and Marrocchi (2017) in NWA 5958 but comparable to the 8 volatility-fractionated patterns found among the 12 aluminum-rich chondrules or "related" objects analyzed in CV chondrites by Zhang et al. (2020). Barred olivine chondrule M8 has the peculiarity to show a flat groundmass REE pattern but a separate mesostasis spot displays a diluted group II pattern (Fig. 1c, 7). The mesostasis in the type II chondrule Mi33, which has only 2 wt% Ca but 6.8 wt% Na (to be compared to the 9.4 wt% and 2.3 wt% (EMPA) averages found in Vigarano by Jacquet, Alard and Gounelle (2012) ), shows a generally flat REE pattern at $\sim 10 \times$ CI with negative anomalies in Eu and Yb (by a factor of 5 and 10, respectively). This is comparable to REE patterns seen in similarly Ca-depleted Sahara 97096 (EH3) chondrule mesostases ( Jacquet et al. (2015a) ), and in Na-rich chondrules in ordinary and Rumuruti chondrites ( Ebert and Bischoff (2016) ), although they also exhibited negative Sm anomalies.

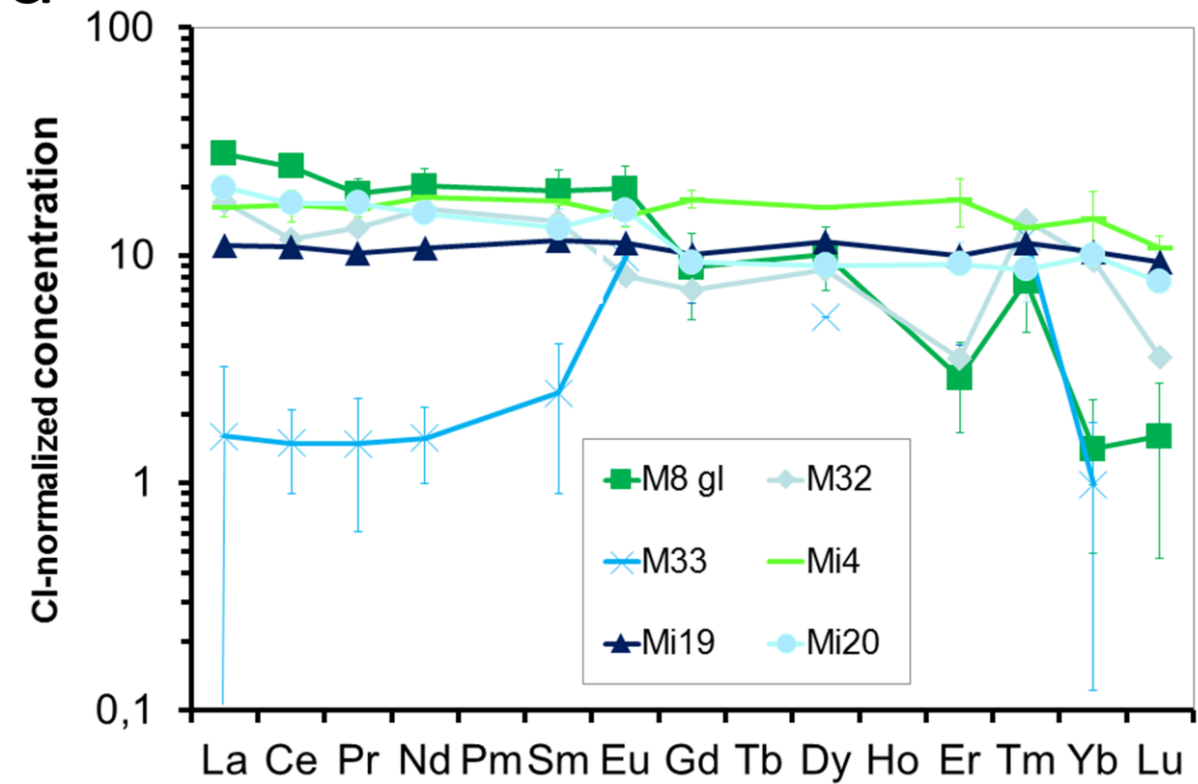
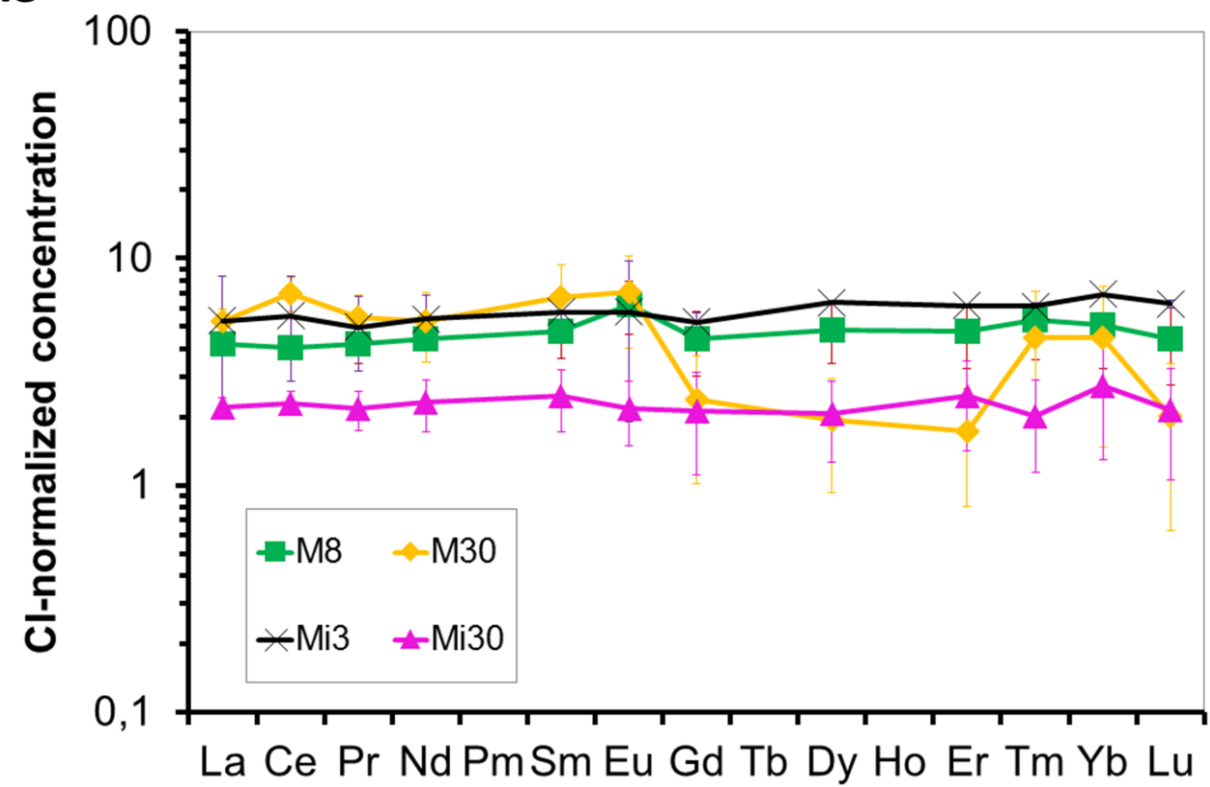

**Figure 7**: REE patterns in mesostases (a) and bulks (b) of mesostasis-rich chondrules. Error bars are one standard deviation.

Let us now turn to the lobate chondrules, the main focus of this paper, and see how the compositions of their lobes correlate. Fig. 8 shows examples of "component *a* vs component *b*" plots, here for Al and Zn. No correlation appears for Al while Zn does show intra-compound correlation.

In order to make this quantitative, I calculated the quadratic average of the intra-chondrule (inter-lobe) standard deviation[1] $\sigma_{intra}$ and normalized to the overall standard deviation of the chondrule lobes $\sigma_{inter}$. That is, I answer the question: "How close are the concentrations of conjoined lobes in a single chondrule compared to what they would be if independently drawn from the total pool of chondrule lobes?". For Al, $\sigma_{intra}/\sigma_{inter} = 0.94$, close to independence. In contrast, $\sigma_{intra}/\sigma_{inter} = 0.48$ for Zn, i.e. the Zn concentrations of two conjoined lobes are closer by a factor of two than in the case of independence. The statistical significance and the relationship with the Pearson correlation coefficient are discussed in appendix A.

The factor determining correlation seems to be volatility. In Fig. 9, I have plotted $\sigma_{intra}/\sigma_{inter}$ as a function of the $T_{50}$ of Lodders (2003). I have also plotted it against compatibility in silicates, as quantified by the Kennedy et al. (1993) olivine/melt partition coefficients D (low-Ca pyroxene shows similar rankings of element compatibilities as the lattice strain model parameters are similar according to Wood and Blundy (2014) ). It is seen that inter-lobe correlation is more related to $T_{50}$ than D. So intra- or inter-chondrule mineral segregation (including the effect of unrepresentative 2D sectioning) does not seem to interfere. Specifically volatile and moderately volatile elements show significant inter-lobe correlation ($\sigma_{intra}/\sigma_{inter} <$ 0.8 for $T_C < 1200$ K) while more refractory elements do not in general. If one single lobate chondrule were particularly predestined to illustrate this, it would be Mi40 where Mi40a is ferromagnesian with weakly fractionated REE while Mi40b has an aluminum-rich chondrule with a diluted modified group II REE pattern (Fig. 5), but both show about average Pb and Zn (0.2-0.3 × CI).

A concern is that analytical (including 2D sectioning) uncertainties may dominate the observed variations, washing out potential intrinsic correlations for refractory elements. I have thus also plotted in Fig. 9 the quadratically averaged normalized analytical error (for those lobes with 2 analytical spots or more). With a median of 0.42, it is usually resolved from 1 but some elements among the most refractory show larger errors within uncertainties of the observed standard deviations, so their deviation from the trend with $T_{50}$ has no intrinsic significance.

---

[1] I refer here to the (Bessel) corrected sample standard deviation (also used for the analytical error), i.e. where, before taking the square root, the sum of the squares of the deviations to the mean is divided by n-1 rather than n, so as to make the expectation value match the actual variance.

The same trends are seen for H3 and CV3 chondrite compound chondrule data from Lux et al. (1981) and Akaki and Nakamura (2005), although limited to elements detectable by electron microprobe (Fig. 9). The Na and K data of Lux et al. (1981) seem to break the trend of increasing correlation with increasing volatility, but this might reflect alkali metasomatism evidenced by mesostasis zoning in unequilibrated ordinary chondrite chondrules (e.g. Grossman et al. (2002) ).

Although Wasson et al. (1995) did not measure bulk compositions of compound chondrule components, they did compare Fe contents in their silicates, as mentioned in the introduction. Fig. 11 shows the strong correlation they found in olivine and low-Ca pyroxene (here lumping together their "sibling" and their "independent" varieties, yielding $\sigma_{intra}/\sigma_{inter} = 0.33$ and 0.27 respectively). This is borne out by my data, also plotted there, even though our general restriction to type I chondrules limits the dynamic range spanned (yielding $\sigma_{intra}/\sigma_{inter} = 0.66$ and 0.95 for olivine and pyroxene, respectively).

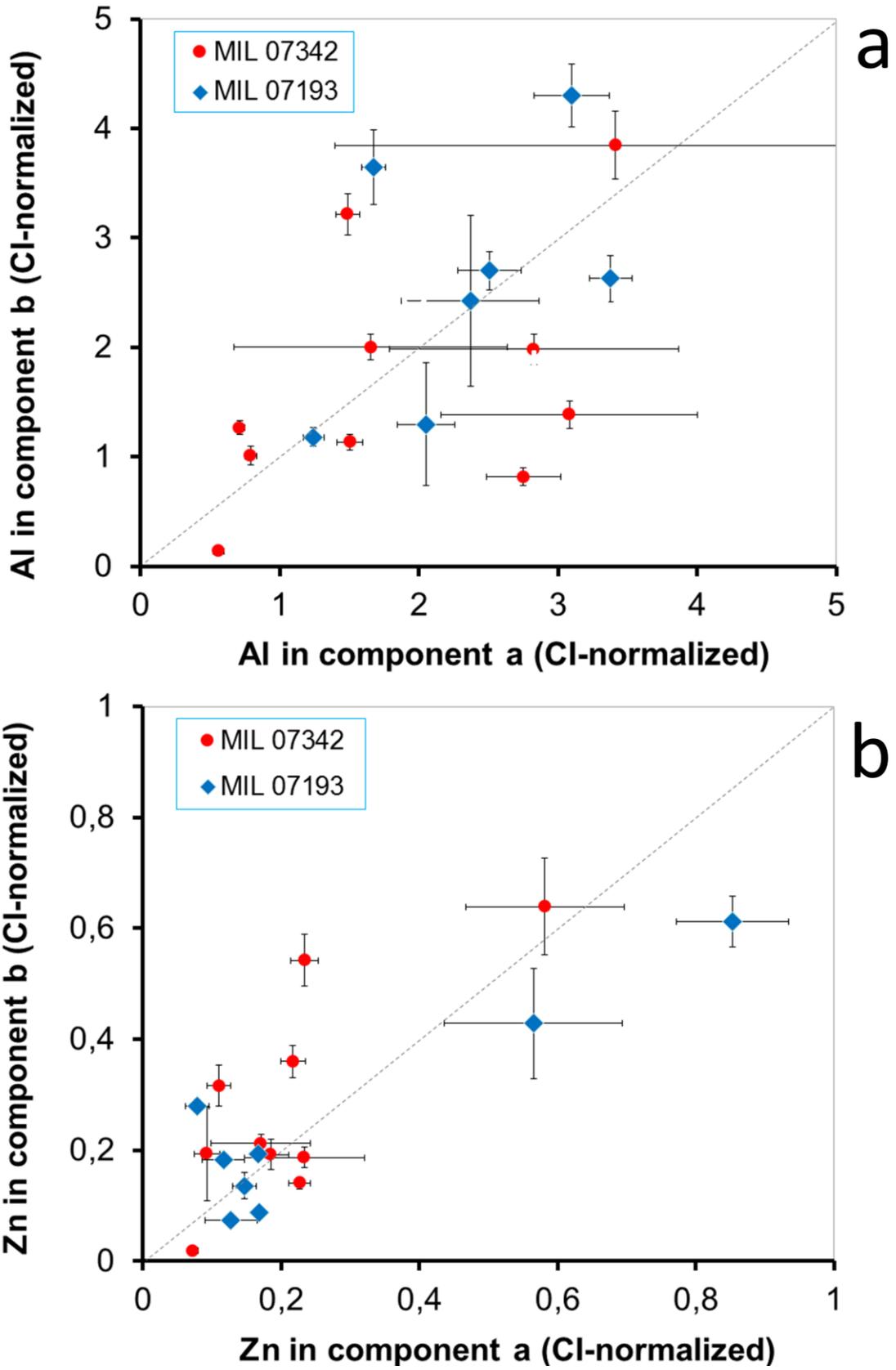

**Figure 8**: Compared Al (a) and Zn (b) contents of two lobes of compound chondrules. Dashed lines mark a 1:1 correlation for reference.

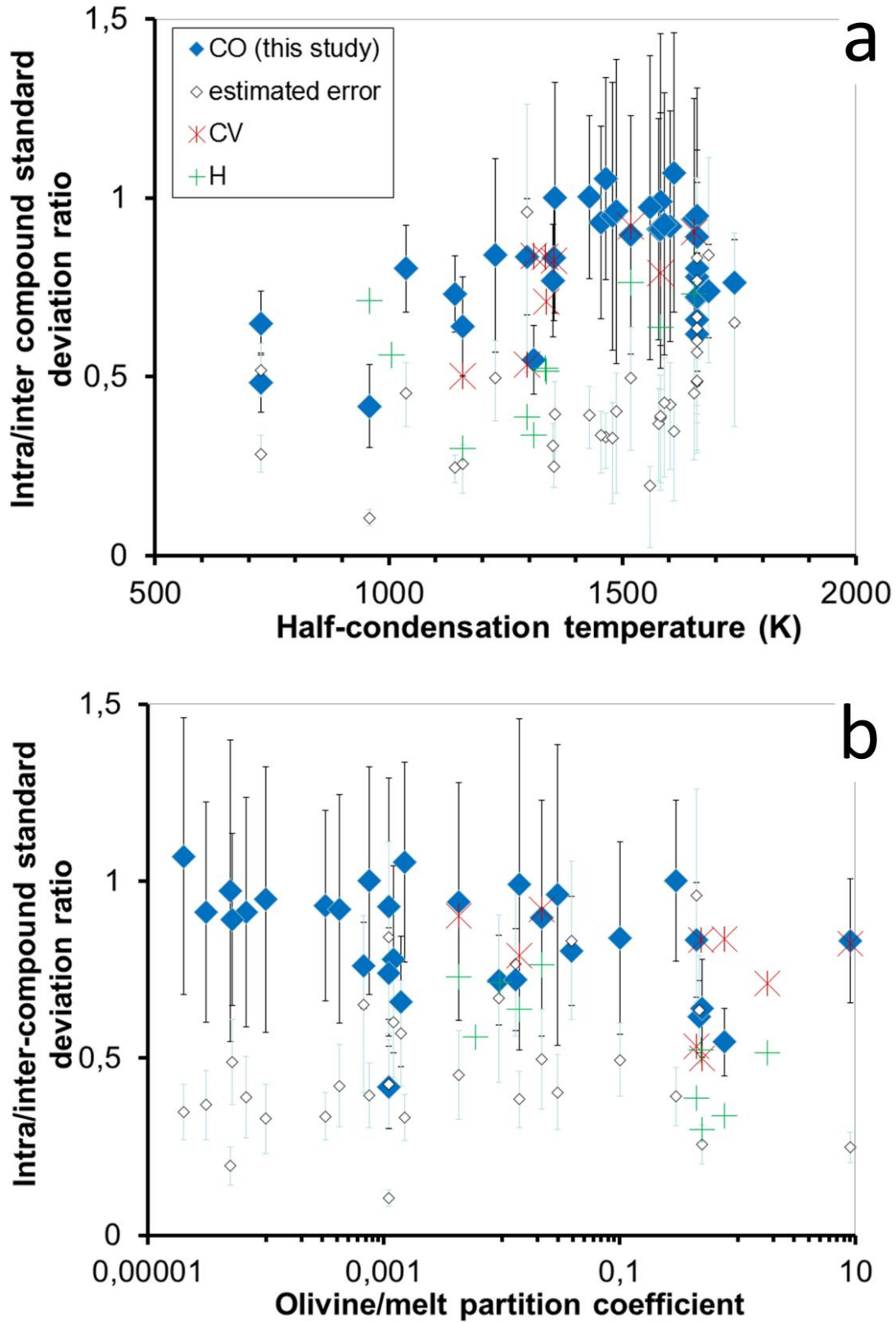

**Figure 9**: Intra-compound standard deviation (quadratically averaged) normalized to the overall standard deviation of the lobe concentrations, plotted as a function of the half-condensation temperature of the elements (a; data from Lodders 2003) and their olivine/melt partition coefficients (b; from run

PO49 of Kennedy et al. (1993) ). I also plot the (quadratically averaged) estimated analytical error (open symbol) normalized to the same overall standard deviation. Error bars are one standard deviation (but do not incorporate the uncertainty of the common normalizing denominators). Data from H3 chondrites (Lux et al. 1981; 16 chondrules) and CV3 chondrites (Akaki and Nakamura 2005; 25 chondrules) are also plotted.

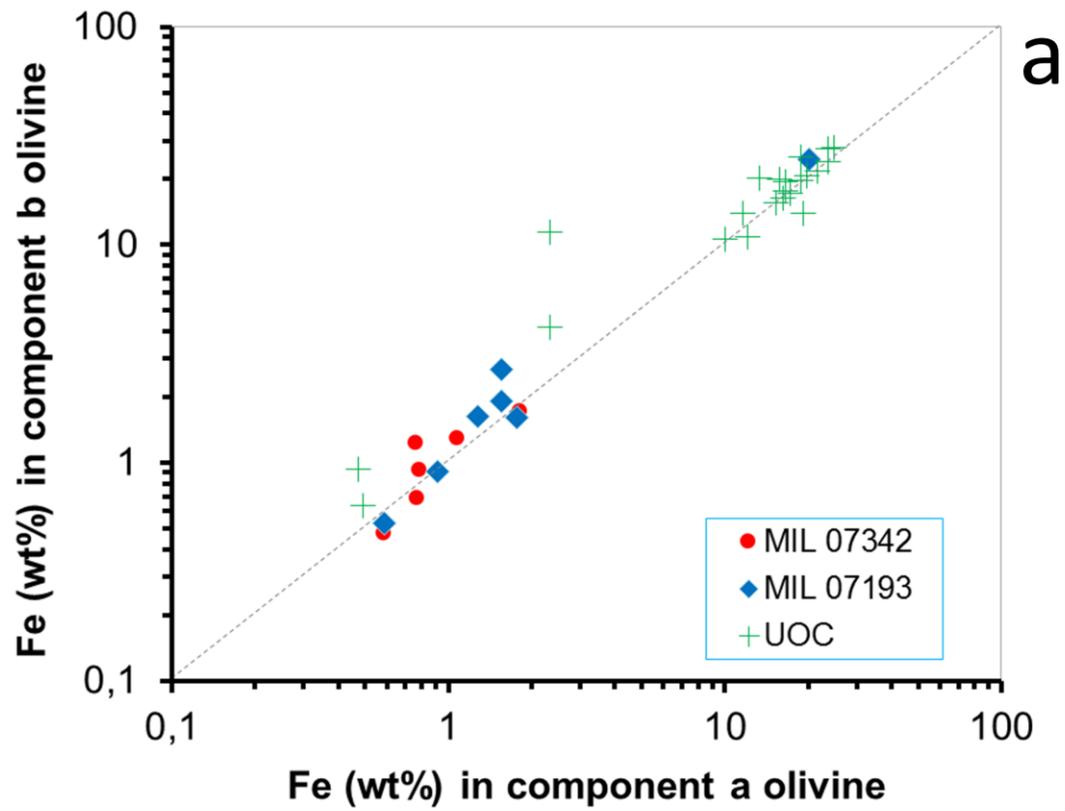

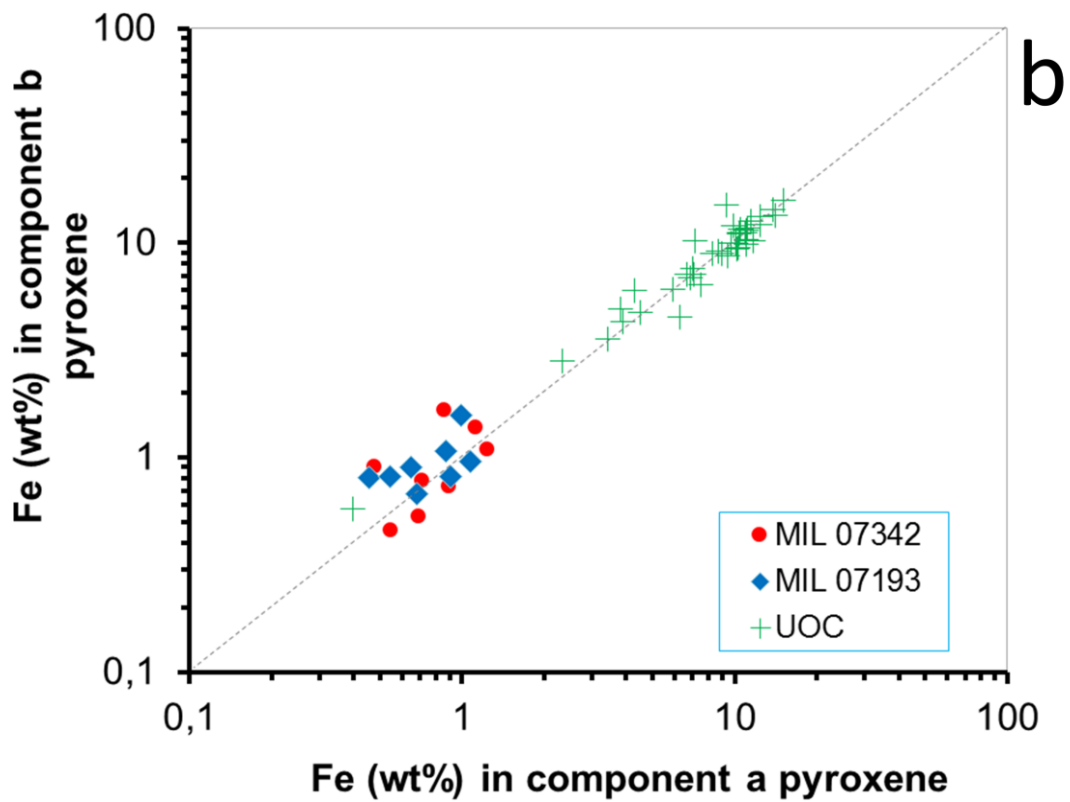

**Figure 10**: Compared Fe concentrations in olivine (a) and low-Ca pyroxene (b) in compound chondrule components. I also plot the unequilibrated ordinary chondrite data of Wasson et al. (1995). Dashed lines correspond to 1:1 correspondence.

## 4. Discussion

### 4.1 Are the lobate chondrules compound?

In this subsection, I will examine whether the lobate chondrules studied here are compound in a genetic sense, i.e. formed by collisions of once-independent droplets.

I first note that the shapes of these objects are different from symmetric rotational shapes such as dumbbells, teardrops or disks (such as those seen among distal impact spherules; Glass and Simonson (2013) )[2]. So if the lobate chondrules are not compound, they must have inherited their shapes from their composite precursors, without having had sufficient time for surface tension to impose overall sphericity. The latter hypothesis was advocated by Rubin and Wasson (2005) for the lobate chondrules in CO chondrites as well as Wasson et al. (1995) for their "independent" compound chondrules (where the primary chondrule was deemed to be a relict directly inherited from the precursor). Rubin and Wasson (2005) did however note that fully molten chondrules should relax very rapidly to sphericity (within a millisecond or so) at liquidus temperature. Hence they preferred lower temperature *and* short-lived melting events to preserve lobate shapes. However, limiting melting to, say, 10-15 % (Rubin and Wasson 2005) would not make viscosity infinite. So let us examine the conditions of relaxation quantitatively.

The timescale of relaxation to sphericity is ( Gross et al. (2013) ):

$$t_{\text{sph}} = \frac{a \eta_{\text{eff}}}{\gamma} \quad (1)$$

with $a$ the droplet radius, $\gamma$ the surface tension of the melt (taken to be 0.36 N/m; e.g. Rubin and Wasson (2005) ) and $\eta_{\text{eff}}$ the effective viscosity of the melt+crystal mix of the chondrule, which can be expressed as a function of the viscosity $\eta_l$ of the melt and the volume fraction of crystals $\phi$ as ( Roscoe (1952) ):

$$\eta_{\text{eff}} = \frac{\eta_l}{(1-\phi)^{5/2}} \quad (2)$$

with $\eta_l$ itself being parameterizable as ( Giordano et al. (2008) ):

---

[2] Appendix B also argues in favor of rapid braking of rotational motions in conditions allowing frequent collisions.

$$\eta_l = \eta_\infty 10^{\frac{B}{T-C}} \quad (3)$$

with $\eta_\infty = 10^{-4.55}$ Pa.s and B and C melt composition-dependent parameters (with the dimensions of temperatures), 5500 and 640 K for my average mesostasis composition, respectively. Thus $\eta_{eff}$ increases with cooling because of (i) the quasi-Arrhenian temperature dependence above (ii) the accumulation of crystals and (iii) the evolution of the residual melt composition toward a more polymerized state.

When will relaxation of an irregularly shaped object to sphericity be prevented? Since $\eta_{eff}$ rapidly increases during cooling, it is sufficient that $t_{sph}$ be initially longer than its own e-folding time $(\mathrm{dln}t_{sph}/\mathrm{d}t)^{-1}$ (essentially $(\mathrm{dln}\eta_l/\mathrm{d}t)^{-1}$), that is $\mathrm{d}t_{sph}/\mathrm{d}t > 1$, so that one can define a critical temperature $T_{crit}$ as:

$$\frac{a\eta_{eff}(T_{crit})}{\gamma} = \frac{(T_{crit}-C)^2}{B\ln(10)\left|\frac{dT}{dt}\right|} \quad (4)$$

with dT/dt the opposite of the cooling rate. Since the temperature dependence of the viscosity on the left-hand-side is strong, the dependence of $T_{crit}$ on the noncompositional parameters is fairly weak. If I adopt $1-\phi = 0.09$ and $a = 0.1$ mm, I obtain $T_{crit} = 1220$ K, 1280 K, 1360 K for cooling rates of 10 K/h, 100 K/h and 1000 K/h, respectively.

It would take *very* rapid cooling rates to raise $T_{crit}$ to the heating peaks invoked by Rubin and Wasson (2005). In order to exceed 1620 K (the onset of low-Ca pyroxene crystallization in the dust/gas = 100 × solar, $10^{-3}$ bar total pressure calculation of Ebel and Grossman (2000)—and thus a generous lower bound of the temperature needed to melt the olivine), we would need cooling rates above 4000 K/s for $\phi = 0.5$. Although the lower limit is comparable to the 3000 K/s predicted from radiative cooling in vacuum at this temperature (e.g. Wasson (1996) ), the latter ignores other heat sources (by radiation or molecular collisions) which would slow it down, depending on the chondrule formation mechanism, as well as empirical evidence of longer heating (e.g. Hewins et al. (2005), Marrocchi et al. (2018) ). It should also be noted that even the advocates of such short flash-heatings invoke repetitions of such events, perhaps a dozen of times ( Baecker et al. (2017), Wasson and Rubin (2003) ), which would in effect allow more time for relaxation to sphericity.

If lobate chondrules did not inherit their shapes from their precursors, then a collisional origin (below $T_{crit}$) becomes unavoidable. In fact, the mineralogical zoning of the individual lobes (olivine in the center, enstatite in the margin; Fig. 3-4) suggests that enstatite grew on them during interaction as independent objects with gaseous SiO (e.g. Barosch et al. (2019), Friend et al. (2016), Libourel et al. (2006)), before they collided (indeed enstatite formation may have started at much higher temperature than $T_{crit} \approx 1300$ K). Had enstatite crystallized only on the present-day lobate objects, olivine necks would have survived connecting their lobes.

Hubbard (2015) envisioned that compound chondrules fused cold (with only a short excursion above 1000 K). This however would not explain the optical continuity of nonporphyritic chondrule (e.g. Wasson et al. 1995), which clearly indicates that the secondary chondrule nucleated on the primary chondrule, and thus must have been a fully molten, undercooled droplet. More generally, it would not explain the correlation of volatile elements among compound chondrule components (Fig. 9a). Indeed, diffusion would not be rapid enough to homogenize the compound: The ratio of the diffusion timescale (through the melt) of an element E $t_{\text{diff,E}} = a^2/D_E$ (with $D_E$ the diffusion coefficient in the melt) to the relaxation timescale is

$$\frac{t_{\text{diff,E}}}{t_{\text{sph}}} = (1-\phi)^{5/2}\frac{a\gamma}{\Omega_E T} = 10^3 \left(\frac{1-\phi}{0.1}\right)^{5/2} \left(\frac{a}{0.1 \text{ mm}}\right) \left(\frac{1300 \text{ K}}{T}\right) \left(\frac{10^{-13} \text{N/K}}{\Omega_E}\right) \quad (5)$$

with $\Omega_E = \eta_l D_E/T$ the "diffusion factor" (Mungall (2002)). In his compilation, Mungall (2002) found it to be around $10^{-13}$ N/K for high-field strength elements, with intermediate field strength elements potentially exceeding this value by up to two orders of magnitude and alkalis scattering somewhat more (but never beyond $10^{-9}$ N/K). So in general $t_{\text{diff,E}} >> t_{\text{sph}}$. This must be why the individual lobes kept their distinctness in refractory elements, which must have essentially behaved in closed-system. Since these would have diffused rapidly in the original independent melts, this distinctness is also diagnostic of the compound nature of a chondrule, even if it has largely relaxed to sphericity.

One reviewer objected that the lobate chondrules show fairly uniform textures, which would favor formation as single objects. However, barring strong compositional differences between the lobe precursors (as in chondrule Mi40; Fig. 5), texture for ferromagnesian chondrules would be largely a fonction of the thermal history (melting for nuclei survival, cooling for crystal growth). Since the lobe components would have formed during the same event, the textural affinities between them appears natural. It may be noted that even in the Rubin and Wasson (2005) scenario of flash heating of preexisting compounds, the explanation for their textural uniformity would be no different.

One prediction of my model is that the oxygen isotopic composition, believed to be buffered by the ambient gas save for relict grains ( e.g. Marrocchi et al. (2018), Ushikubo et al. (2012)

), should be the same in the different component of a lobate chondrule. The first measurements of this kind by Akaki and Nakamura (2005), on a Cameca 6f probe, could not have enough precision to resolve the range spanned by type I chondrules in carbonaceous chondrites; most chondrules should anyway belong to the $\Delta^{17}O \approx -5$ ‰ mode of Ushikubo et al. (2012) as in most carbonaceous chondrites except CR (e.g. Jacquet et al., in press). I nonetheless note that type I chondrule N6-10 analyzed by Jacquet and Marrocchi (2017) in NWA 5958, was compound (see Fig. 2a of Jacquet et al. (2016) ), and had its main component and all its four analyzed adhesions (analyses N6-10@3,4,5,6) comprised between -2.9 ± 0.2 and -2.0 ± 0.2 ‰ in $\Delta^{17}O$. This is statistically significant given the relative rarity of $\Delta^{17}O > -3$ ‰ compositions among type I chondrules (4 out of 13 type I chondrules analyzed by Jacquet and Marrocchi 2017).

The correlations of the volatile and moderately volatile elements[3] mentioned above, although statistically significant, are not strong (see e.g. Fig. A1 in the appendix) but they actually need not be in the emerging scenario of a single buffering gaseous environment. Indeed, for fixed partial pressures of the species of interest, the bulk concentration of volatile elements will still depend on the melt/crystal ratio in the chondrule and the melt concentrations themselves will depend on the activity coefficients and thence the abundance of other major elements (such as refractory ones; see e.g. Mathieu et al. (2011) for the case of sodium). The intra-compound correlation of the FeO contents of the silicates is much stronger (Fig. 10), likely because FeO is more unequivocally related to the oxygen fugacity of the medium (e.g. Schrader et al. 2013; Tenner et al. 2015). I note that after attempting to distinguish "sibling" from "independent" compound chondrules (only the former being assumed to have formed by collisions during chondrule formation), Wasson et al. (1995) remarked that the latter also exhibited such a correlation (see their Fig. 6). Although they attributed this to broad temporal proximity of the formation of the components in a secularly cooling nebula, the latter astronomical picture has given way to that of a generally cold protoplanetary disk, where chondrule-forming events are spatially and temporally localized ( Jacquet et al. (2015b) ). So I conclude that the Wasson et al. (1995) distinction between sibling and independent compound chondrules is *in fine* arbitrary and that all compound chondrules[4] formed by collisions during chondrule-melting events.

---

[3] The conditions of chondrule formation were surely different from those of the Lodders (2003) calculation but the latter should still give a reasonable proxy of the relative volatility of the elements.
[4] Enveloping compound chondrules also described by Wasson et al. (1995) are discussed in the following section.

## 4.2 Implications on compound chondrule statistics

The argument on the brevity of $t_{sph}$ near liquidus temperatures mentioned in the preceding section means that essentially any chondrule appreciably deviating from sphericity must be compound (in the genetic sense of resulting from a merger between two or more liquid droplets), unless rotation or postsolidification deformation can be invoked. Rubin and Wasson (2005) found that 60-75 % of the chondrules in CO3 chondrites Allan Hills 77307, Colony and Yamato 81020 had aspect ratios above 1.2. Nelson and Rubin (2002) reported 60 % of chondrules with aspect ratios above 1.2 in Semarkona (LL3.0) from optical microscopy measurement. Of course, 2D measurements underestimate the departure from sphericity of chondrules. After correcting for postaccretional deformation, Charles et al. (2018) still found that 100 out of 109 chondrules of CR2 chondrite NWA 801 examined in X-ray tomography had aspect ratios above 1.1, and 52 above 1.3. Twelve of the 20 Allende (CV3) chondrules studied by Tsuchiyama et al. (2003) with the same technique had aspect ratios exceeding 1.1. So it would seem that a majority of chondrules suffered mergers. Of course, in many cases, relaxation to sphericity has made the lobes no longer readily recognizable, so those that are only represent the tip of the iceberg, as entertained by Alexander and Ebel (2012). One example of a serendipitous "cryptocompound" chondrule, part of my previous trace element study in Bishunpur (Jacquet et al. 2015b), only revealed as such in an X-ray map, is depicted in Fig. 11. Further examples may be the two L/LL3.05 chondrite chondrules studied by Pape et al. (2021), although these author attributed their textural and mesostasis compositional zoning to two successive melting events: indeed, their nonspherical shape and/or zoning rather evoke (enveloping) compound chondrules formed by collisions, whether they were sectioned close to the interface (Ciesla et al. 2004), or really feature one chondrule engulfed in another[5]. In the present study, the REE pattern discrepancy between the mesostasis and groundmass measurement in barred olivine chondrule M8 (Fig. 7a,b) may indicate a largely resorbed compound (with a PO chondrule?)—indeed, as shown in the previous subsection, the diffusion timescale of refractory elements should be slower than relaxation—, possibly also betrayed by difference in olivine textures (Fig. 1c). Petrographically, chondrule Mi19 (Fig. 1d) may also evoke a compound between a BO (top), whose olivine shell would have been largely dissolved, and a PP chondrule (bottom), but trace element evidence is here lacking (the two mesostasis LA-ICP-MS spots straddle the putative boundary and show indistinguishable flat REE patterns).

---

[5] The evolution of melt composition through mixing would account for the compositionally distinct overgrowths in the pyroxene crystals.

Collisions may have involved droplets much smaller than common chondrules. This may be the case for microchondrules seen to adhere on many ordinary chondrite chondrules (e.g. Dobrica and Brearley (2016) ). Jacquet et al. (2015b) even interpreted coarse-grained igneous rims (from which part of the "enveloping compound chondrules" of Wasson et al. (1995) may only differ by a matter of degree) as the result of the accretion of microdroplets at the end of chondrule cooling. These interpretations are at variance with the commonly cited works of Rubin and Wasson (1987) and Krot and Wasson (1995) who ascribe igneous rims to later remelting of fine-grained accretionary rims or chondrule margins, as well as Krot et al. (1997) and Bigolski et al. (2016) who consider adhering microchondrules to be *forming* by budding (through some unspecified physics). In this case of separate heatings, however, the correlation of FeO/MgO ratios of silicates between igneous rims and chondrule interiors (Krot and Wasson 1995) would be difficult to understand (Jacquet et al. 2015b). There are, it is true, fayalitic microchondrules trapped in fine-grained rims around type I chondrules, but these—which are much more ferroan than typical type II chondrules—might at least in part result from parent body alteration of more magnesian microdroplets, as suggested by occasional ferroan halos around the latter with the reverse relationship never being seen (Bigolski et al. 2016). It thus appears simpler to invoke a single heating/coagulation period. It would also more directly account for the correlation between the prevalence of igneous rims and the paucity of nonporphyritic textures (favored by nucleation-inducing impinging microdroplets) seen by Rubin (2010) as evidence for remelting of dust rims for the former. The microchondrules might have been generated by chondrule-chondrule collisions ( Dobrica and Brearley (2016) ), similar to the origin envisioned by Jacquet et al. (in press) for isolated olivine grains, some of which subsequently formed compounds. Collisions, or inertial or surface tension effects may also have ejected isolated metal grains, some of which were apparently recaptured at chondrule margins, especially for metal-armoured CR chondrite chondrules ( Jacquet et al. (2013) ).

Given that small droplets easily outnumber the bigger ones (even for a tiny mass fraction), "normal"-sized chondrules may well (nearly) all be compound in the widest genetic sense of having undergone a merger with at least one droplet, since collisions with the larger ones affected at least a tenth or so of them (from the lower limit given by earlier literature estimates; e.g. Gooding and Keil 1981; Wasson et al. 1995; Akaki and Nakamura 2005). Now it may be useful to restrict the empirical definition of compound chondrules by setting some arbitrary threshold in aspect ratio or other sphericity metrics, effectively a limit on the size ratio of the merged droplets as well as their subsequent relaxation degree (that is, the temperature of the collision). The number of the larger merged droplets per chondrule cannot be too large. In fact, the compositional diversity of chondrules precludes the coalescence of more than a dozen similar-size droplets per chondrule ( Hezel and Palme (2007) ). With one "compoundness" criterion or another, tomographic studies could compare more objectively the abundances of compound chondrules across chondrite groups. They should further characterize the possibility of postaccretional deformation, e.g. shear or lineation (Charles et al. 2018). The CO chondrites may e.g. still offer the highest abundances, which may be consistent with other petrographic evidence of higher solid densities in the chondrule-forming regions of carbonaceous chondrites ( Rubin (2010) ).

However one defines compound chondrules, I thus infer that the average number of mergers $N_{coag}$ per chondrule with similar-size droplets during its partially molten time, was about unity, give or take one order of magnitude, the lower limit relying on literature estimates of the (recognizable) compound chondrule fraction and the upper limit being set by the constraint of compositional diversity (Hezel and Palme 2007). This crude $N_{coag} \approx 10^{0\pm1}$ estimate will be sufficient for the remainder of the discussion.

While I thus suggest that compound chondrule abundances may have been underestimated by about one order of magnitude in the literature, the *nonporphyritic* compound chondrule abundances should be more robust, and they might be less liable to erasure by surface tension effects because of crystallization before (primary) or during (secondary) collision ( Jacquet (2014) ) unless their silicates became unstable in the composite melt. Ciesla et al. (2004) estimate a compound fraction of 20 % for them. So it is conceivable that contrary to earlier claims, nonporphyritic chondrules have a *lower* compound fraction than their porphyritic counterparts, although that would not necessarily imply a less collisional environment, as the primary chondrules may have had to experience a first collision to induce nucleation prior to the collision with the secondary ( Arakawa and Nakamoto (2016) ). Still, it does seem that nonporphyritic chondrules formed in events distinct from their porphyritic counterparts. Indeed, there is a definite "endogamy" among nonporphyritic chondrules. Wasson et al. (1995) listed 55 nonporphyritic-nonporphyritic chondrules, 10 nonporphyritic-porphyritic ones and 14 porphyritic-nonporphyritic ones (in the order primary-secondary), so a nonporphyritic chondrule (primary or secondary) was at least 4 times more likely to pair with a nonporphyritic than a porphyritic "colleague". As an extreme, an Allende "macrochondrule" depicted in Fig. 5 of Weyrauch and Bischoff (2012) comprised no less than 16 conjoined BO chondrules and no other textural type. So most nonporphyritic chondrules probably represent distinct thermal histories, probably under longer peak heating (e.g. Hewins and Fox (2004) ) and/or more rapid cooling rates ( Radomsky and Hewins (1990) ).

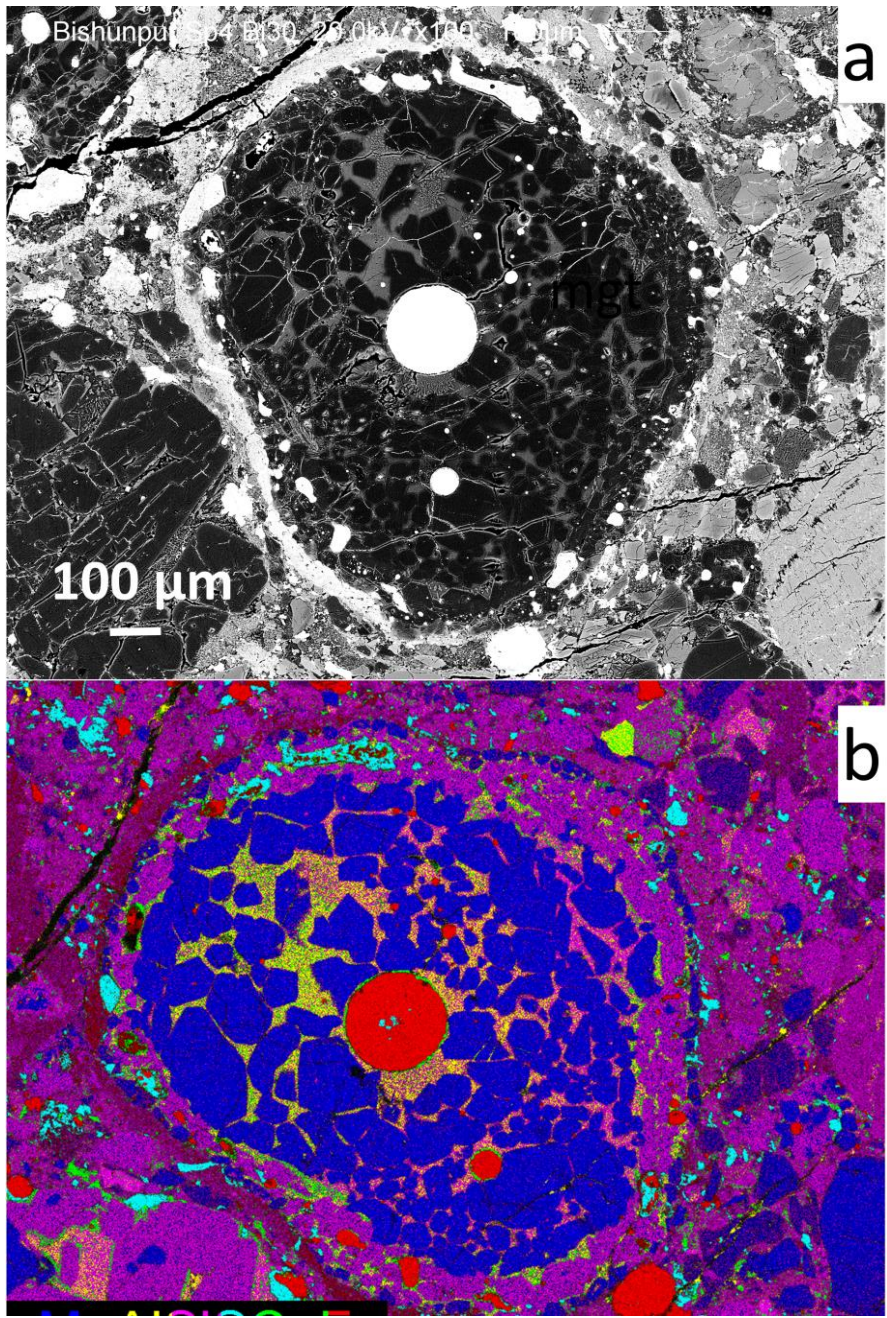

**Figure 11**: Chondrule Bi30 in Bishunpur (LL3.15), analyzed by Jacquet et al. (2015b). Panel (a) is a BSE image and panel (b) is a composite X-ray map with the same color coding as above

(mind the ~30° counterclockwise rotation). While panel (a) suggests little more than a regular porphyritic olivine texture, panel (b) reveals that chondrule Bi30 is actually compound, and a primary chondrule, with a prominent ~200 µm diameter Fe-Ni nodule, may be defined (on the upper left), possibly because the olivine palisade (the coarser-grained outer part) allowed it some rigidity. Its Al-rich mesostasis has not efficiently mixed with the Al-poorer one of the secondary. Enstatite occurs only as a rim interspersed with Fe-Ni and sulfide grains around the whole compound chondrule, and it itself surrounded by a discontinuous layer of olivine.

## 4.3 An impact origin for chondrules?

If I denote by t the partially molten time of chondrules, $n_c$ the number density of particles (assumed to be similar in size and density $\rho_s$ to the reference target chondrule) and $\Delta v$ the relative velocity, I have:

$$N_{\text{coag}} = s n_c 4\pi a^2 \Delta v\, t \qquad (6)$$

with $s$ ($\leq 1$) a sticking probability. This can be solved for the condensable density $\rho_c = n_c 4\pi \rho_s a^3/3$ as:

$$\rho_c = \frac{N_{\text{coag}} \rho_s a}{3 \Delta v\, t} = 3 \times 10^{-4} \text{kg/m}^3 N_{\text{coag}} s^{-1} \left(\frac{a}{0.1 \text{ mm}}\right)\left(\frac{0.1 \text{ m/s}}{\Delta v}\right)\left(\frac{1\text{h}}{t}\right) \qquad (7)$$

Here I have uncritically (for the time being) adopted the numbers of Alexander and Ebel (2012) for the normalizations, but in yielding densities much in excess of the *total* density (gas included) expected in a Minimum Mass Solar Nebula ( Hayashi (1981) ), this suffices to show that compound chondrules challenge expectations from "nebular" scenarios. It is but natural that planetary scenarios, in particular impact scenarios, whether on solid targets (Johnson et al. 2015) or on magma oceans (Asphaug et al. 2011) are entertained on this basis, reinforced by the present work.

However, compound structures, although not unheard of among terrestrial impact spherules, seem to account for less than a percent of them ( Glass (1974), Glass and Simonson (2013) ). Their frequency might increase with proximity to the source crater, where glass fragments possibly related to particle-particle collisions (if not to thermal stress upon entry into the ocean; Glass and Simonson 2013) abound. Yet so does the percentage of unmelted ejecta (Glass and Simonson 2013), limiting the analogy with chondrule-dominated chondrites (Taylor et al. 1983). Conversely, rotational shapes (dumbbells, disks, teardrops) widespread among terrestrial microtektites and microkrystites are not commonly seen in chondrules, as mentioned in section 4.1.

The primitive composition of chondrules is also a problem. Lichtenberg et al. (2018) found that a magma ocean should rapidly locally homogenize and globally differentiate. So, absent trends of igneous fractionations among chondrules (e.g. Grossman (1988), Wood (1996) ), the "splashing" scenario would need to target planetesimals on the verge of differentiation. This would however do little to avoid homogenization upon the impact. Taylor et al. (1983) already noted that impact melts had the compositions of target rocks rather than individual minerals. This is true for microtektites as well, whose compositions largely overlap with their macroscopic tektite counterparts (Glass and Simonson 2013), although the microtektites may extend to lower (but not higher) silica contents, perhaps owing to different depths of origin in the source crater depending on the landing location. Schlieren in some Australasian microtektites (hereafter "microaustralasites") within ~1000 km from Indochina do suggest incomplete mixing at submillimeter scale (e.g. Folco, Perchiazzi, D'Orazio, Frezzotti, Glass and Rochette (2010) ) but the siliceous composition of the terrestrial continental crust involves much slower diffusion than the mafic chemistries of ferromagnesian chondrules. If I adopt the average microaustralasite composition of Koeberl (1990), I obtain viscosity parameters of $B = 9847$ K and $C = 453$ K (see section 4.1), hence a difference of 3 to 10 orders of magnitude in viscosity (and hence diffusivity) compared to bulk chondrule compositions ($B = 5167$ K, $C = 592$ K for our averaged main components), between 2000 and 1000 K. The mafic impact spherules found at the Lonar crater match the basaltic target bulk composition, rather than forming a tie line between feldspar and pyroxene, if the homogenization scale had been smaller than crystal size, ~50 μm judging from the supplements of Ray et al. 2017. More precisely, the standard deviations for different elements measured with 50 μm LA-ICP-MS spots within a factor of two of those of the target rocks analyzed at 20-50 g scales ( Osae et al. (2005), Ray et al. (2017) ). This, of course, excepting those elements appreciably affected by contamination from the (enstatite chondrite; Ray et al. 2017) impactor (those with impactor/target concentration ratio above 2, e.g. siderophile elements).

When identifying impact spherules (whether melted droplets or, a fortiori, condensates) to target rock compositions plus possible (whole-rock) impactor admixture, I should in fact allow for open-system behavior, as seen for microaustralasites where alkalis are depleted with increasing craterocentric distance ( Folco, Glass, D'Orazio and Rochette (2010) ). Open system behavior was more pronounced in the peculiar CB and CH nonporphyritic "chondrules" ( Krot et al. (2001) ), which are believed to have formed in impact plumes (e.g. Krot et al. (2005) ). The skeletal olivine and cryptocrystalline ones are respectively enriched and depleted in refractory elements, with both exhibiting depletions in moderately volatile elements, but their REE patterns are quite flat (less than ~10 % relative anomalies).

Indeed flat REE patterns seem characteristic of meteorite parent bodies. In their bulk meteorite analyses, Dauphas and Pourmand (2015) found $Tm/Tm^*$ ranging between 0.94 (aubrites) and 1.13 (Allende). This is a lower variability than our chondrules, with a relative standard deviation of $Tm/Tm^*$ of compound chondrule main components of 17 % and deviations of a factor of a few in some other chondrules (up to ~4 here or in Jacquet and Marrocchi 2017). More generally, refractory element ratios vary more in chondrules than in bulk chondrites, as can be seen in Fig. 6 where the error bars attributed to the CO bulk correspond to an unweighted standard deviation of CO, CM and CV group averages, as a proxy for the planetesimal variability in the formation reservoir of CO chondrules (see also Ebel et al. (2016), Lux et al. (1981) ). Now one could envision that bulk chondrules represent a wider, now unsampled set of planetesimal compositions. However, we would then expect refractory elements in compound chondrule lobes to correlate (at least as well as moderately volatiles), as they would share the same precursor parent bodies, so that the intra-chondrule-forming-event variability would be dominated by the inter-chondrule-forming-event variability. This is contrary to observations (Fig. 9).

Clearly, the subdued group II-like patterns must reflect the presence of large calcium-aluminum-rich inclusions (CAIs), not diluted by any significant averaging, in the precursors of these chondrules (Misawa and Nakamura 1988; Jones and Schilk 2009; Jacquet and Marrocchi 2017; Zhang et al. 2020). This is supported by their higher frequency in mesostasis (and thence refractory element)-rich chondrules. This also accounts for correlation between Ti concentrations and Ti isotopic anomalies ( Gerber et al. (2017) ), which incidentally is especially inconsistent with the splashing scenario, as even the heating prior to magma ocean formation should have erased those anomalies (and their relationship to refractory components) at the mm scale. Not all fractionated REE patterns necessarily correspond to CAI precursors; that of chondrule Mi33, with prominent negative Eu and Yb anomalies (Fig. 2), and possibly many Na-rich chondrules of Ebert et al. (2018) may reflect oldhamite formation and removal in the precursor (accounting for the Ca depletion), similarly to that advocated by Jacquet et al. (2015a) for enstatite chondrite chondrules. The presence of silica in Mi33 and the sodium enrichment of its mesostasis (as well as that of the Na-rich chondrules of Ebert et al. 2018) is also reminiscent of enstatite chondrite chondrule parageneses ( Jacquet et al. (2018), Piani et al. (2016) ).

Thus, thoroughly homogenized chondritic protoliths (with or without preexisting chondrule) cannot match the wide range of chondrule composition. A chondritic protolith is then conceivable only if the homogenization was limited to short spatial scales. Since the contiguous sub-mm-size inclusions in a chondrite are independent of each other, the standard deviation of element concentrations should roughly decrease as the mixing lengthscale to the -3/2 power, once that lengthscale goes beyond a millimeter or so. As an interesting illustration, the thirty-eight ~0.6 g (~0.2 cm$^3$) chips of Allende analyzed by Stracke et al. (2012) never deviate by more than 27 % in Tm/Tm* from the average (with a relative standard deviation of 11 %), already at variance with the factor-of-a-few anomalies shown by chondrules with volatility-fractionated REE. Clearly averaging cannot extend beyond a few millimeters, even in so coarse-grained a chondrite as Allende, to reproduce chondrule compositional variability. So the compositions of chondrules are best understood as deriving from free-floating precursors of similar sub-mm size, and which themselves resulted from the stochastic aggregation of more or less refractory grains ( Jacquet (2014) ). This is more in line with a "nebular" setting for chondrule formation.

### 4.4 Or a collisional nebular setting?

However, if we go back to "nebular" scenarios, the question as to the solid enrichments of chondrule-forming regions resurfaces at once. Setting compound chondrule frequencies aside for a moment, how are we to explain solid/gas ratios ε of order unity suggested by type I chondrule olivine fayalite content ( (e.g. Schrader et al. (2013), Tenner et al. (2015) )?

Sekiya and Nakamura (1996) advocated settling of solids to the disk midplane. Given the likely presence of turbulence, the resulting enhancement factor would be $(1+S_z)^{1/2}$ where $S_z$ is the settling parameter $S_z = St/\delta_z$, with $\delta_z$ the non-dimensionalized turbulent diffusivity (e.g. Jacquet (2013), Jacquet, Gounelle and Fromang (2012) ) and St the Stokes number, which in Epstein regime can be expressed as:

$$St = \Omega \sqrt{\frac{\pi}{8} \frac{\rho_s a}{\rho_g c_s}} \quad (8)$$

with $\rho_s$ and *a* the internal density and size of the particles, respectively, $\Omega$ the Keplerian angular velocity, and $\rho_g$ and $c_s$ the gas density and isothermal sound speed, respectively. In a turbulent disk where $\delta_z$ is of order the Shakura and Sunyaev (1973) α parameter, $S_z$ should not exceed unity by more than two orders of magnitude ( Jacquet, Gounelle and Fromang (2012) ), so the maximum enhancement would be still one order of magnitude off that required for even type I chondrules. This prompted Hubbard et al. (2018) to question the validity of chondrule concentration estimates from cosmochemical evidence.

However, the paradigm of protoplanetary disk accretion may be changing. Beyond the inner region where the magnetorotational instability (MRI) may be sustained by thermal ionization alone, nonideal magnetohydrodynamical effect are likely to suppress the MRI over the whole thickness of the disk (e.g. Bai (2017), Turner et al. (2015) ) and accretion through this "dead zone" may proceed largely because of the torques induced by disk winds, in the absence of widespread turbulence. If the disk midplane is laminar, streaming instabilities and/or vertical shear instabilities would regulate the disk thickness to about 1 % of the pressure scale height *H* of the disk (e.g. Bai and Stone (2010), Schäfer et al. (2020) ), which would generically reproduce the ε ~ 1 indicated by type I chondrule mineral composition. This, however, assumes that residual turbulence does not interfere, which puts strong constraints (upper limits) on allowable $\delta_z$. These might though be alleviated over the evolution of the disk, as gas densities drop, which may be allowed by the apparent 1-2 Ma delay in chondrule formation relative to CAIs (e.g. Nagashima et al. (2018), Villeneuve et al. (2009) ). Of course, it may be that the chondrule-forming process itself (or some prelude to it) somehow locally concentrated solids.

Still, the densities indicated by equation (7) are challenging even assuming ε ~ 1. However, note that we have not yet discussed parameters, in particular the relative velocity in chondrule-forming regions, which must be considered an unknown on a par with the density. I may rewrite equation (6) in terms of ε:

$$N_{\text{coag}} = 3\sqrt{\frac{\pi}{8}}\, s\, \varepsilon\, \Omega t\, \frac{\Delta v}{c_s St} \quad (9)$$

For ε ~ 1 and $t$ of order a few hours, give or take one order of magnitude (that is $\Omega t$ of order $10^{-3\pm1}$) it is clear that, for $N_{\text{coag}}$ ~$10^{0\pm1}$ $\Delta v/(c_s St)$ has to be $\gg$ 1. For reference, in a minimum mass solar nebula model (e.g. Hayashi (1981) ), at an heliocentric distance R = 3 AU, $c_s St$ would be ~0.1 m/s at the midplane. Although the implied relative velocities would likely exceed fragmentation velocities of solid aggregates (1-10 m/s; e.g. Blum (2018) ), partially molten chondrules should be able to withstand collision velocities up to 1 km/s (and form compounds) because of their viscosity ( Arakawa and Nakamoto (2019) ).

Yet it should be noted that such relative velocities cannot be expected from "normal" disc turbulence. This has been already claimed by Sekiya and Nakamura (1996) but can be demonstrated explicitly here. If I adopt equation 3 of Jacquet (2014), and take typical size differences to be of order $a$, such turbulence and differential drift would yield:

$$\frac{\Delta v}{c_s St} = \max\left(\frac{1}{\sqrt{1+S_z}}, \left|\frac{\partial \ln P}{\partial \ln R}\right|\frac{H}{R}, \min\left(\sqrt{\delta_z}\text{Re}^{1/4}, \sqrt{\frac{3}{S_z}}\right)\right) (10)$$

with P the pressure[6] and Re the Reynolds number. Since any significant degree of settling requires $S_z > 1$, and since H/R $\ll$ 1 (i.e. the disk is geometrically thin), it is clear that $\Delta v/(c_s St)$ must be < 1, regardless of the particulars of the adopted disk model. In other words, larger relative velocities would correspond to a broader dust subdisk, thwarting any concentration by settling.

Thus, the chondrule formation process cannot merely operate in (or create) a region of high solid concentration, but needs to boost relative velocities. Such high velocities and/or high chondrule concentration should however be transient to avoid wholesale shattering of chondrules after glass transition, where the critical fragmentation velocity would drop below ~150 m/s ( Arakawa and Nakamoto (2019), Ueda et al. (2001) ). At face value, the lightning model (e.g. Desch and Cuzzi (2000), Johansen and Okuzumi (2018) ) does not appear to give such a boost. A candidate that naturally comes to mind would be the shock wave model (e.g.

---

[6] I have decomposed the pressure gradient term on the left in its radial and vertical component, the typical value of the latter (assumed to be a Gaussian of standard deviation H around the midplane) being dictated by the thickness of the dust subdisk H/$(1+S_z)^{1/2}$ (e.g. Jacquet et al. (2013) ).

Desch et al. (2005) ), because of the differential slow-down of variously sized particles relative to the post-shock gas (e.g. Arakawa and Nakamoto (2019), Ciesla (2006), Nakamoto and Miura (2004) ). However it may be just *too* efficient in developing collisions. Jacquet and Thompson (2014) found that even before melting, the chondrule precursors would be destroyed by mutual collisions and sandblasting by finer dust even assuming moderate enhancements of solids above solar. While droplet recoagulation downstream would be conceivable, the resulting chondrule compositions would represent averages of the local reservoir, stumbling on the same criticism raised against planetary scenarios in the preceding subsection.

## 5. Conclusion

In this work, I have shown that lobate chondrules in CO chondrites were most likely compound objects, formed by the coalescence of two or more partially molten droplets. Those behaved in closed system for refractory elements but exchanged more volatile elements with the same gaseous environment, as often petrographically manifested by the enstatite-rich margins around each lobe. In fact, since surface tension rapidly imposes spherical shapes at high temperatures, nonspherical chondrules, which are a majority in different chondrite groups, are probably more or less collapsed compounds formed below about 1000 °C. Given that most chondrules across chondrite groups appreciably deviate from sphericity, the compound fraction for porphyritic chondrules may have been hitherto underestimated by about one order of magnitude.

It would certainly be useful to measure more accurately the abundance of compound chondrules, picking some threshold of departure from sphericity, e.g. by X-ray tomography, across chondrite groups so as to establish any difference between chondrite clans (e.g. between carbonaceous and non-carbonaceous chondrites). It would be important to clearly distinguish what departure from sphericity can be due to postaccretional deformation (e.g. upon impact). Further chemical analyses (*in situ*, or even more exhaustive analyses on chondrules extracted from disaggregated chondrites) would likewise be desirable, with perhaps a wider array of moderately volatile elements than measured here.

Still, the conclusion that collisions must have been very common in chondrule-forming regions already has strong implications on their nature. While an impact origin may first come to mind, this is little consistent with the compositional variability of chondrules, which cannot represent mixtures of known whole-rock target and impactor compositions, or the lack of intra-compound refractory element correlation (in this study or literature data from CV or ordinary chondrites) that would be implied by a common parent body for the precursors. Chondrule chemistries are most compatible with the melting of sub-millimeter-size aggregates of protoplanetary disk dust. However, a nebular setting for chondrule formation is also challenging. Enhanced solid/gas ratios of order unity, as suggested by the FeO contents of type I chondrules, *might* be conceivable in a sufficiently dead zone of the protoplanetary disk, but the chondrule-forming process would also need to boost relative velocities much above background turbulence to satisfy compound chondrule statistics. At any rate, the frequent collisions inferred for chondrule-forming regions must be considered on a par with compositional or thermal constraints in the evaluation of chondrule formation models.

*Acknowledgments*: I thank the three referees Dr Alan Rubin, Denton Ebel and Tomoki Nakamura, for allowing to clarify the motivation of the project and the logic of the discussion, as well as completing the bibliographical references. I thank Roger Hewins for a read of a first draft. I am grateful to Bill Glass for his readiness to answer my questions on microtektites. I thank the Meteorite Working Group at the Johnson Space Center for the loan of the MIL 07342 and MIL 07193 samples. US Antarctic meteorite samples are recovered by the Antarctic Search for Meteorites (ANSMET) program which has been funded by NSF and NASA, and characterized and curated by the Department of Mineral Sciences of the Smithsonian Institution and Astromaterials Curation Office at NASA Johnson Space Center. This research was funded by the Programme National de Planétologie (project "L'héritage nébulaire des chondres", 2016). Last and not least, I am grateful to Christine and Pierre de Naurois, and Georges Illy and Michèle Ridoux for hosting me during my numerous analyses in Montpellier. This paper is dedicated to the memory of my aunt Marylène Joly.

# Appendix A: Intra-compound correlation coefficient

In the main text, I have chosen $\sigma_{intra}/\sigma_{inter}$ as a metric of the correlation between lobes of compound chondrule. It may be worth commenting how it relates to the more standard Pearson correlation coefficient r between lobe *a* and *b* concentrations of a given element (e.g. the biplots of Fig. 8), which however is not symmetric with respect to the change of the *a* and *b* labels in individual chondrules (e.g. if sectioning biases misrepresent which one is the bigger lobe). Let us call A and B the random variables corresponding to the concentration of an element of interest in lobes *a* and *b* of a compound chondrule, respectively. Then, by definition:

$$r = \frac{E(AB) - E(A)E(B)}{\sigma_A \sigma_B}$$

with E the expectation value and $\sigma_{A,B}$ the standard deviation of A, B. Note that r is not squared, so potentially take any value, positive or negative, between -1 and 1. For example, r = 0.13 for Al and r = 0.77 for Zn in the plots of Fig. 8. Assuming normal distributions, independent variables should yield r<0.4 with a probability of 95 % for n=18 chondrules; I can thus take 0.4 as a rough threshold for the statistical *significance* of a positive correlation (but not for its *strength*, of course), so the Zn concentrations of two lobes in a given lobate chondrule *are* significantly correlated, unlike Al.

I have:

$$\left(\frac{\sigma_{intra}}{\sigma_{inter}}\right)^2 = \frac{E[(A-B)^2]/2}{[E(A^2)+E(B^2)]/2 - [(E(A)+E(B))/2]^2}$$

Which yields:

$$\left(\frac{\sigma_{intra}}{\sigma_{inter}}\right)^2 = 1 + \frac{E(A-B)^2/2 - 2\sigma_A\sigma_B r}{E(A-B)^2/2 + \sigma_A^2 + \sigma_B^2} = 1 - \frac{2\sigma_A\sigma_B}{\sigma_A^2 + \sigma_B^2}r$$

Where the second equality assumes E(A)=E(B). If we can further assume $\sigma_A = \sigma_B$ (which however may be inaccurate as larger instrumental errors attend smaller LA-ICP-MS spots), a simple relationship results:

$$\frac{\sigma_{intra}}{\sigma_{inter}} = \sqrt{1-r}$$

So r>0.4 should roughly correspond to $\sigma_{intra}/\sigma_{inter}$ < 0.77 as observed for volatile and moderately volatile elements here. It should thus come as no surprise that plots of r (Fig. A1) show the same trends as the plots on $\sigma_{intra}/\sigma_{inter}$. It is noteworthy, though, that the r of the refractory elements, though small, generally remain on the positive side, possibly in part due to common levels of dilution after recondensation of more volatile elements.

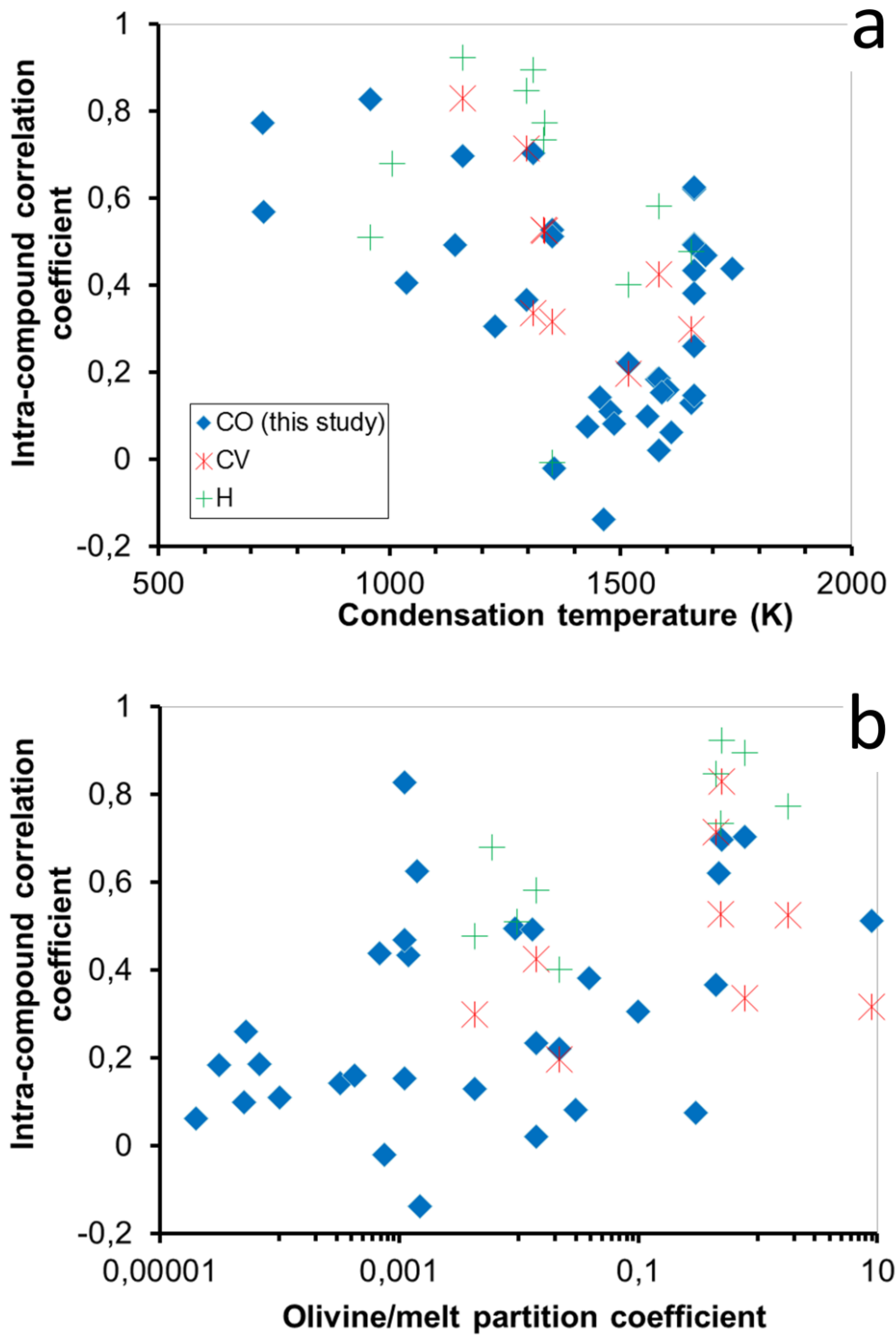

**Figure A1**: Inter-lobe correlation coefficient of element concentrations plotted as a function of the half-condensation temperature of the elements (a; data from Lodders 2003) and their olivine/melt partition coefficients (b; from run PO49 of Kennedy et al. (1993) ). Data from H3 chondrites (Lux et al. 1981; 16 chondrules) and CV3 chondrites (Akaki and Nakamura 2005; 25 chondrules) are also plotted (with the primary component equated to the main component, as nearly always the case; Wasson et al. 1995).

# Appendix B: Rotation braking

In the main text, I have argued against rotation being a significant factor to explain chondrule nonsphericity from morphological comparisons (with *bona fide* rotational shapes like disks, dumbbells and teardrops). Now an initial rotation of a chondrule precursor is certainly conceivable, but on being partly molten, it would rapidly (on timescale $t_{sph}$) relax to a symmetric (relative to the center of gravity), if not yet spherical (because of the centrifugal force), shape. The interaction with the gas would then cause braking of the rotational motion, on a timescale comparable to the linear stopping time *τ = St/Ω* in Epstein conditions (see e.g. appendix of Epstein (1924) for a spherical surface, bearing in mind that the moment of inertia tensor of an homogeneous sphere is the scalar *8πρ$_s$a$^5$/15*). Then, from equation (9):

$$\frac{\tau}{t} = 3\sqrt{\frac{\pi}{8}} \frac{s\varepsilon}{N_{\text{coag}}} \frac{\Delta v}{c_s}$$

So long relative velocities are subsonic[7], it appears that this is << 1 if $N_{\text{coag}}$ is of order ~1 (or greater). In other words, a substantial compound chondrule fraction (in fact, even the lower limits suggested by past literature, depending on *Δv/c$_s$*) implies (indirectly) efficient braking, and thus a subordinate role of rotation in the observed chondrule shapes, consistent with morphological observations.

---

[7] At face value from the equation, ε should not be too much greater than unity but then collisions with other droplets would contribute to the braking, leading to τ~ρ$_s$a/max(ρc$_s$, ρ$_c$Δv) essentially cancelling the ε *Δv/c$_s$* factor in the right-hand-side.